\documentclass[prd,amsmath,amssymb,floatfix,superscriptaddress,twocolumn]{revtex4-2}

\usepackage{ulem}

\include{pack}

\def\aap{A\&A}
\def\apj{ApJ}

\def\mnras{MNRAS}

\def\aj{AJ}

\def\physrep{Phys. Rep.}

\def\apjs{ApJS}

\usepackage{times,graphics}
\usepackage{makecell}
\usepackage{diagbox}

\usepackage{graphicx}
\usepackage{dcolumn}
\usepackage{bm}
\usepackage{natbib}
\usepackage{graphicx}
\usepackage{epsfig}
\usepackage{epstopdf}
\usepackage{multirow}
\usepackage{amsmath}
\usepackage{lipsum}
\usepackage{amssymb}
\usepackage{color}
\usepackage{dcolumn}
\usepackage{bm}
\usepackage{natbib}
\usepackage{tabu}
\usepackage{times,graphics}
\usepackage{makecell}
\usepackage{diagbox}

\RequirePackage{mathptmx}      
\RequirePackage{flushend}

\begin{document}
	
\title{Using newest VLT-KMOS HII Galaxies and other cosmic tracers to test the $\Lambda$CDM tension}

\author{Ahmad Mehrabi }
\affiliation{Department of Physics, Bu-Ali Sina University, Hamedan, 65178, 016016, Iran}
\email{Mehrabi@ipm.ir}

\author{Spyros Basilakos}
\affiliation{Academy of Athens, Research Center for Astronomy $\&$ Applied Mathematics, Soranou Efessiou 4, 11-527, Athens, Greece }
\affiliation{National Observatory of Athens, Lofos Nymfon, 11852 Athens, Greece}
\email{svasil@academyofathens.gr}

\author{Pavlina Tsiapi}
\affiliation{National Technical University of Athens, 9 Iroon Polytechniou St., 15780, Greece}

\author{Manolis Plionis}
\affiliation{National Observatory of Athens, Lofos Nymfon, 11852 Athens, Greece}
\affiliation{Physics Department, Aristotle University of Thessaloniki, Thessaloniki 54124, Greece}

\author{Roberto Terlevich}
\affiliation{Instituto Nacional de Astrof\'{i}sica, Optica y Electr\'{o}nica,Tonantzintla,C.P. 72840, Puebla, M\'{e}xico}
\affiliation{Institute of Astronomy, University of Cambridge, Cambridge, CB3 0HA, UK}

\author{Elena Terlevich}
\affiliation{Instituto Nacional de Astrof\'{i}sica, Optica y Electr\'{o}nica,Tonantzintla,C.P. 72840, Puebla, M\'{e}xico}

\author{Ana Luisa Gonzalez Moran}
\affiliation{Instituto Nacional de Astrof\'{i}sica, Optica y Electr\'{o}nica,Tonantzintla,C.P. 72840, Puebla, M\'{e}xico}

\author{Ricardo Chavez}
\affiliation{CONACyT-Instituto de Radioastronom\'{i}a y Astrof\'{i}sica, UNAM, Campus Morelia, C.P. 58089, Morelia, M\'{e}xico}

\author{Fabio Bresolin}
\affiliation{Institute for Astronomy, University of Hawaii, 2680 Woodlawn Drive, 96822 Honolulu,HI USA}

\author{David Fernandez Arenas}
\affiliation{Instituto Nacional de Astrof\'{i}sica, Optica y Electr\'{o}nica,Tonantzintla,C.P. 72840, Puebla, M\'{e}xico}
\affiliation{Kavli Institute for Astronomy and Astrophysics, Peking University, Beijing 100871, China}

\author{Eduardo Telles}
\affiliation{Observatorio Nacional, Rua Jos\'{e} Cristino 77, 20921-400 Rio de Janeiro, Brasil}

\begin{abstract}
	We place novel constraints on the cosmokinetic parameters
	by using a joint analysis of the newest VLT-KMOS HII galaxies (HIIG)
	with the Supernovae Type Ia (SNIa) Pantheon sample.
	We combine the latter data sets in order 
	to reconstruct, in a model 
	independent way, the Hubble diagram to as high redshifts as possible. 
	Using a Gaussian process we derive the 
	basic cosmokinetic parameters and compare them with 
	those of $\Lambda$CDM. In the case of SNIa we find that the
	extracted values of the cosmokinetic parameters 
	are in agreement with the predictions of $\Lambda$CDM model. 
	Combining SNIa with high redshift tracers of the Hubble relation, namely 
	HIIG data
	we obtain consistent results with those based on $\Lambda$CDM
	as far as the present values of the cosmokinetic parameters
	are concerned, but find significant deviations in the evolution of the 
	cosmokinetic parameters with respect to the expectations of the concordance $\Lambda$CDM model.

\end{abstract}

\maketitle


\section{Introduction}
The discovery of the accelerated expansion of the Universe from 
Supernovae type Ia (SNIa) 
data \citep{Riess1998,Perlmutter1999} has opened a new window for research 
in Cosmology. Indeed, the 
analysis of various cosmological probes, including those of 
Cosmic Microwave Background (CMB) \citep{Komatsu2011,Ade:2015yua,Aghanim:2018eyx}, Baryon Acoustic Oscillation (BAO) \citep{Eisenstein:2005su,Percival2010,Blake:2011rj,Reid:2012sw,Abbott:2017wcz,Alam:2016hwk,Gil-Marin:2018cgo}, and 
cosmic chronometers \cite{Farooq:2016zwm} have provided  
the general framework of cosmic expansion, namely the universe 
is in the phase of acceleration at late times. 
Despite the latter confirmation, 
the nature of cosmic acceleration remains a 
mystery, hence a lot of effort has been put in by 
cosmologists 
over the last two decades in order to provide a viable 
explanation concerning the underlying mechanism which is responsible 
for this phenomenon. 

Within the framework of homogeneous and isotropic Universe, 
the corresponding accelerated expansion 
can be described by 
considering either a new form of matter 
with negative pressure \citep{Weinberg:1989,Peebles:2002gy,Copeland:2006wr,Chiba:2009nh,Amendola:2010,Mehrabi:2018dru,Mehrabi:2018oke} or a modification 
of gravity \cite[$f(R)$, $f(T)$ theories etc,][]{Schmidt:1990gb,Magnano:1993bd,Dobado:1994qp,Capozziello:2003tk,Carroll:2003wy}. 
Among the large family of cosmological models 
the  present  accelerating  phase  of  the  universe  is  quite
well  described  in  the  context  of  general  relativity together with a cosmological constant -- the so called
$\Lambda$CDM model. This model is spatially flat, with   
cold dark matter (CDM) and baryonic matter coexisting with the
cosmological constant.
However, the $\Lambda$CDM model suffers from  well known theoretical 
problems, namely the coincidence problem and the 
expected value of the vacuum energy density 
\citep{1989RvMP...61....1W,2003PhR...380..235P,perivolaropoulos2008puzzles,padilla2015lectures}. 

Although the $\Lambda$CDM model is 
consistent with the majority of cosmological 
data \citep{Aghanim:2018eyx}, the model seems 
to currently be in tension with some recent 
measurements \citep{Verde_2019,Sola:2017znb,DIVALENTINO2021102605,DIVALENTINO2021102604,Perivolaropoulos:2021jda},
associated with the Hubble constant $H_0$ and 
the present value of the mass variance at 8$h^{-1}$Mpc, namely $\sigma_8$. 
Also  \cite{Lusso:2019akb} performing  
a combined analysis of SNIa, quasars, and gamma-ray bursts (GRBs), 
found a $\sim 4\sigma$ tension between the 
best fit cosmographic parameters with respect to those of 
$\Lambda$CDM
(see also \cite{Risaliti:2018reu}).
Recently, \cite{Lusso_2020} continued this study by
using the same notations but a relatively larger sample 
of quasars. They have shown a strong 
deviation from $\Lambda$CDM at high redshifts. 
As expected, in light
of the aforementioned results, an intense debate is taking place 
in the literature and our work attempts to contribute to this debate.

In our work, for the first time, we use 
our new set of high spectral
resolution observations of high-z HIIG, obtained with VLT-KMOS 
\citep{Gonz_lez_Mor_n_2021}
along with available HIIG data \citep{Gonzalez-Moran:2019uij,Terlevich:2015toa}
and combine them with the Type Ia Supernovae (SNIa) data
from the Pantheon sample \citep{Scolnic:2017caz} 
in order to reconstruct the Hubble diagram 
and thus to place constraints on the main cosmokinetic parameters
({\em deceleration} and {\em jerk}), as well as to check for 
deviations from the predictions of  the $\Lambda$CDM model.
Specifically, in this paper we focus on a model-independent 
parameterization of the Hubble diagram using the Gaussian process, 
and investigate its performance against the latest SNIa+HIIG
Hubble diagram data. It is crucial to note that we need to introduce 
a kernel function with a set of hyper-parameters which can be optimized 
in order to fit the data. The reader may find more details of  
model-independent methods in  
\citep{Liao:2019qoc,Zhang:2018gjb,Gomez-Valent:2018hwc,Melia:2018tzi,Mehrabi_2020}. 

The structure of the current article is as follows. 
In section  \ref{sec:back}, we briefly review the 
basic cosmological background equations and present  
the observational data used.
In section \ref{sec:gau}, we present the main properties of 
the Gaussian process together with the corresponding 
kernel functions, while in section \ref{sec:res} we discuss our results. 
Finally, we draw our conclusions in section \ref{sec:con}.

\section{Background cosmology and data set}\label{sec:back}
Considering a spatially 
flat FRW cosmology and assuming that the fluid components 
do not interact with each other, the evolution of the Hubble parameter 
is given by
\begin{equation}\label{eq:hub-back}
H^2 (z)= H_0^2\left[\Omega_{m0}(1+z)^3 + \Omega_{r0}(1+z)^4 + \Omega_{x}(z)\right],
\end{equation}
where $H_0$ is the Hubble constant, 
$\Omega_{m0}$ and $\Omega_{r0}$ are the density parameters of 
matter and radiation at the present time. 
The parameter $\Omega_{x}(z)$ can be seen as a 
general function that describes the dark energy component. 
There are a lot of options for the 
dark energy component 
(see the introduction for some references), but the simplest case is 
a constant density parameter: 
$\Omega_\Lambda(z) = \Omega_{\Lambda 0} = 1-\Omega_{m0}-\Omega_{r0}$, namely 
the well known 
$\Lambda$CDM model, where 
the Hubble parameter is written as 
\begin{equation}\label{eq:hub-lcdm}
H^2 (z)= H_0^2\left[\Omega_{m0}(1+z)^3 + \Omega_{r0}(1+z)^4 + 1-\Omega_{m0}-\Omega_{r0}\right].
\end{equation}
Of course, it is straightforward to compute the luminosity distance,
\begin{equation}\label{eq:lum-dis1}
D_L(z) = (1+z)D(z),
\end{equation}
where
\begin{equation}\label{eq:com}
D(z) = \int_0^z \frac{dx}{H(x)},
\end{equation}
is the comoving distance, while the luminosity distance usually is related 
with the distance modulus,
\begin{equation}\label{eq:mod-dis}
\mu(z) = 5\log_{10}D_L(z) + 25.
\end{equation}
where the distance $D_L$ is given in Mpc.
Having a sample of standard candles in the universe and measuring 
their distance moduli, it is straightforward to compute the Hubble parameter from
\begin{equation} \label{eq:H_from_com}
H(z) = \left[D'(z)\right]^{-1},
\end{equation}
where prime denotes derivative with respect to redshift and $D'(z) \ne 0$.
Therefore, given the luminosity distance data, one can reconstruct 
the comoving distance and then find the corresponding Hubble parameter.  
The step that follows is to derive the 
cosmokinetic parameters, namely
\begin{eqnarray}\label{eq:q-j-h}
&q(z)& = (1+z)\frac{H'(z)}{H(z)} - 1,\\
&j(z)& = (1+z)^2 \left[\frac{H''(z)}{H(z)}+(\frac{H'(z)}{H(z)})^2\right]
- 2(1+z)\frac{H'(z)}{H(z)}  + 1,
\end{eqnarray}  
where $q(z)$ and $j(z)$ are the 
deceleration and jerk parameters, respectively. 
These parameters provide a useful parametrization, in
these kind of studies, to test 
the performance of a given cosmological model against the 
observational data.

Below we briefly present the type of observational data used in 
reconstructing the Hubble diagram. 

\begin{figure}
	\centering
	\includegraphics[width=9.5 cm]{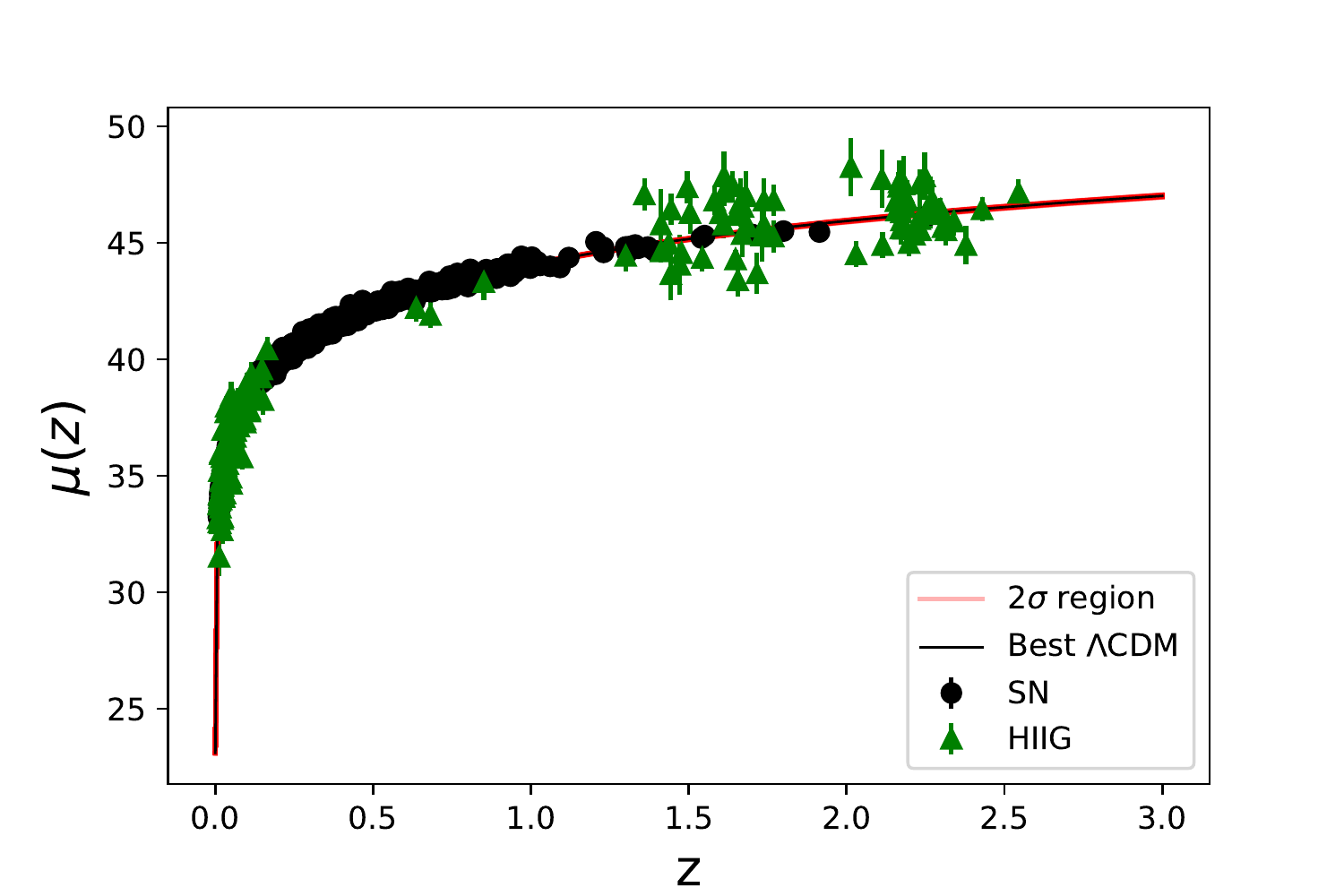}
	\caption{The distance moduli as a function of redshift of the SNIa and
		HIIG are shown as points with uncertainties. The continuous curve is the
		best-fit $\Lambda$CDM Hubble function.}
	\label{fig:data}
\end{figure}

\begin{figure}
	\centering
	\includegraphics[width=8.7 cm]{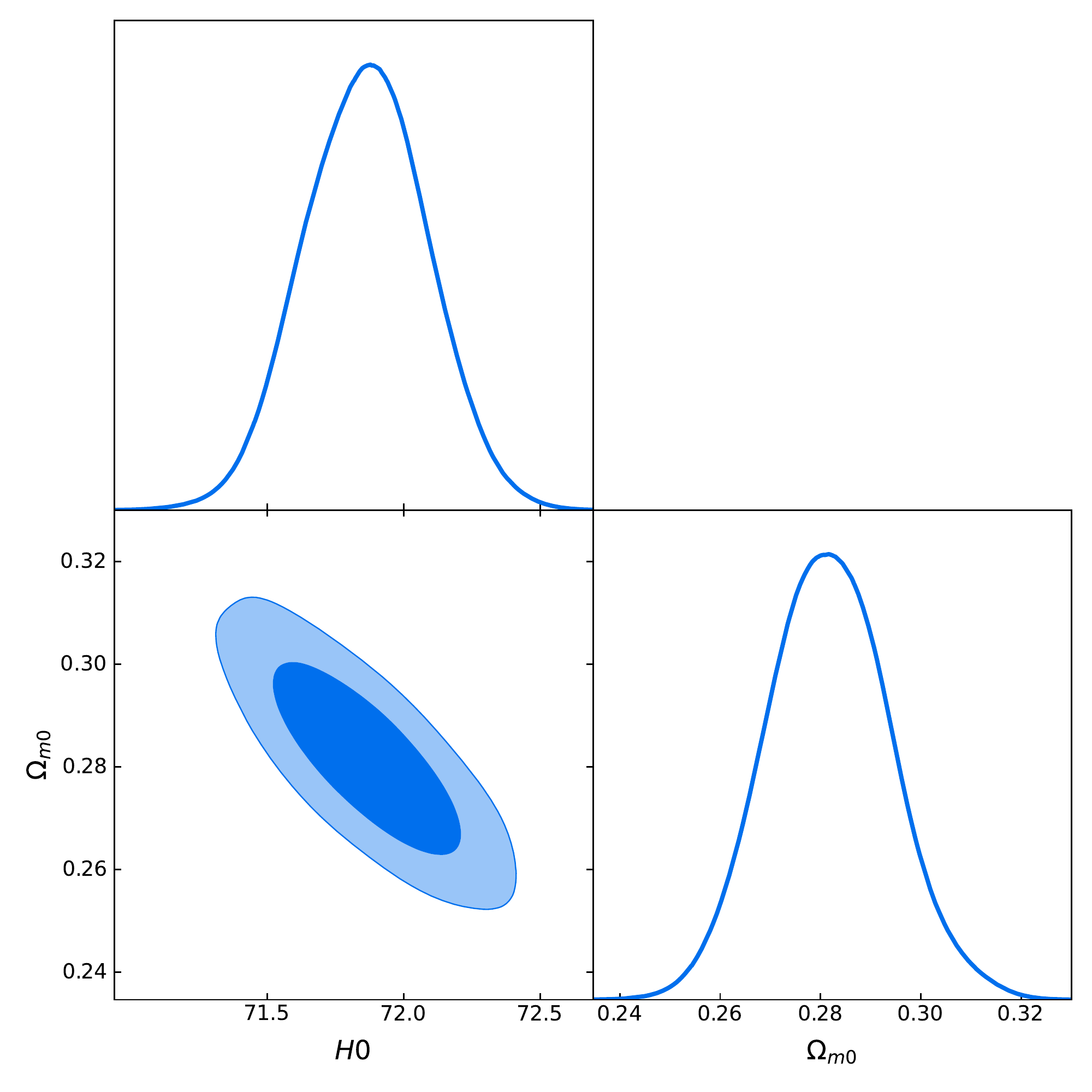}
	\caption{1 and 2 $\sigma$ confidence levels in the $H_0-\Omega_{m0}$ plane and their posterior distributions for the $\Lambda$CDM model.}
	\label{fig:conf}
\end{figure}

\begin{itemize}
	
	\item HIIG data: In a large sequence of articles, \citep{Gonz_lez_Mor_n_2021,Gonzalez-Moran:2019uij,Fernandez-Arenas:2017isq,Chavez:2016epc,Terlevich:2015toa,Plionis:2009wp,Plionis:2011jj,Plionis:2009up,Melnick:1999qb} 
	HIIGs have been proposed as alternative tracers of the cosmic expansion
	extending the Hubble diagram on epochs beyond the range of available
	SNIa data. We use the HIIG sample as provided by \cite{Gonz_lez_Mor_n_2021} (for a full description of the data see \citep{Gonz_lez_Mor_n_2021}).
	This data set includes
	181 entries, which can be split in the local sample (107
	objects: $z\le 0.16$) and a high redshift sample which contains 
	29 KMOS, 15 MOSFIRE, 6 XShooter and 24 literature objects \citep{Gonz_lez_Mor_n_2021}. The redshift span of the HIIG data is: $0.088 \le z \le 2.5$.

	\item SNIa data: In our analysis we utilize the data of the
	Pantheon SNIa sample (considering only statistical uncertainties) \citep{Scolnic:2017caz},
	which combines the SNIa sources of the 
	PanSTARRS1 (PS1) Medium Deep Survey with a large number of other
	surveys, to compare the two distinct and independent tracers of the
	Hubble relation.
	
\end{itemize}
Before reconstructing the Hubble parameter, we perform a Bayesian 
inference considering the $\Lambda$CDM model to find the 
cosmokinetic parameters at the present time. In Figure (\ref{fig:data}), 
we present the observational data and on top we plot 
the predictions of $\Lambda$CDM model by using the best fit values, namely 
$\Omega_{m0}=0.282\pm 0.012$ and $H_{0}=71.86\pm0.22$\footnote{For the 
	concordance $\Lambda$CDM model we have $q_{\Lambda}(z)=\frac{3}{2}\Omega_{m}(z)-1$ 
	and $j_{\Lambda}(z)=1$, where $\Omega_{m}(z)=\Omega_{m0}(1+z)^{3}/E^{2}_{\Lambda}(z)$
	with 
	$E^{2}_{\Lambda}(z)=H^{2}_{\Lambda}(z)/H^{2}_{0}=\Omega_{m0}(1+z)^{3}+1-\Omega_{m0}$.}.
The comparison indicates that there is a 
small difference between data 
and model 
at high redshifts. 
Additionally, in Figure (\ref{fig:conf})
we present the contour plot in $H_{0}-\Omega_{m0}$ plane. 
Notice that we do not consider the radiation 
term in Eq.(\ref{eq:hub-back}) since we are well inside 
the matter/dark energy dominated eras.

\section{Gaussian process as a model independent method}\label{sec:gau}
In general, based on the luminosity distance, 
there are two main avenues in order to study the Hubble diagram, aiming  
to understand the expansion rate of the universe.  
The first choice is to impose a cosmological model 
and through standard lines to extract the corresponding 
form of the luminosity distance. 
Subsequently, the model is fitted to the
data in order to put constraints on the 
corresponding free parameters. Obviously, this approach  
is a model-dependent method, since  
different models provide different functional forms 
for the luminosity distance.
The second choice is to use a model independent method in building
the Hubble diagram via the observational data
\citep{Liao:2019qoc,Zhang:2018gjb,Gomez-Valent:2018hwc,Melia:2018tzi}, hence 
we do not need to impose a particular model as an underlying cosmology.
In this context,
one of the model independent methods which is widely used in this kind
of studies is the Gaussian process (GP),
and indeed in the present article we attempt to test the performance of 
GP against the available Hubble diagram data. 

Now let us briefly present the basic steps of the method.
The GP is a sequence of Gaussian random variables (RV), which 
can be modeled with a multivariate Gaussian distribution. 


Generally, for a given data set $M$
\begin{equation}\label{eq:data-set}
M =\{(x_i,y_i)|i=1,..,n\},
\end{equation}
our target is to build a function which can reproduce the data 
in a model independent way, assuming that  
the data points can be modeled with a GP, 
$y\sim GP(\mu(x),K(x_i,x_j))$, where $\mu(x)$ provides the mean value at each
point and $K$ is the so called GP kernel which indicates the
covariance function between different points. In this case, the diagonal
(off-diagonal) terms of the kernel give the uncertainty at each point
(correlation between different points).
Now, in order to build a continuous function we need to compute the prediction of
GP at a set of arbitrary points ($x^{\star}$) rather than at the observational points
($x$).  In this context, it is an easy task to compute the mean and the covariance 
function at these new points according to \cite{10.5555/1162254}:
\begin{eqnarray}\label{eq:GP}
\mu^{\star} &=& K(x,x^{\star})[K(x,x^{\star})+C_D]^{-1}Y\\
\Sigma^{\star} &=& K(x^{\star},x^{\star}) - K(x^{\star},x)[K(x,x^{\star})+C_D]^{-1}K(x,x^{\star}), 
\end{eqnarray}
where $C_D$ is the covariance matrix of the data, $Y$ 
is the column vector of observation $y_i$, while a zero mean prior has
been considered in deriving the above equations. 
Having obtained these quantities, one can generate many 
functions at $(x^{\star})$ using the expression
$$f(x^{\star})\sim N(\mu^\star,\Sigma^\star),$$
where $N(\mu^\star,\Sigma^\star)$ is a multivariate Gaussian distribution.

Regarding the kernel function, there are a variety of options 
depending of course on the characteristics of the data. 
The best known kernel is the squared exponential kernel 
\begin{equation}\label{eq:gau-ker}
K(x_i,x_j) = \sigma_f^2\exp{\left(-\frac{(x_i-x_j)^2}{2l^2}\right)},  
\end{equation}
where $\sigma_f^2$ and $l$ are two hyper-parameters of 
the kernel. Since the results might depend on the selected kernel, it 
is common procedure to consider several kernels. 
In our analysis, we utilize the Matren $\nu=7/2$ and $\nu=9/2$ kernels 
along with equation (\ref{eq:gau-ker}) as detailed in 
\cite{Mehrabi_2020}. 

It is interesting to mention that the GP method provides
the reconstructed function $f(x)$ as well as the corresponding derivatives.
Indeed, since the derivative of a GP function is another GP, the 
corresponding derivative with respect to variable $x$ 
is given by
$$f'(x)\sim GP\left(\mu'(x),\frac{\partial^2K }{\partial x_i\partial {x_j}}\right)$$
Concerning the second and third derivatives we refer the reader for more details to \cite{GP_book,Seikel:2012uu}.


To conclude this section let us summarize the steps that we need to follow 
in order to reconstruct the Hubble diagram and the corresponding 
cosmokinetic parameters.\\
\begin{enumerate}
	\item We convert the distance modulus to luminosity distance and finally to comoving distance, \\ 
	\item we obtain the GP of $D(z)$ using different kernels and different data sets, by employing the {\it{scikit-learn}} library \citep{scikit-learn}, \\
	\item for $D'(z)=0$, we see from equation (\ref{eq:H_from_com}) 
	that $H(z)$ goes to infinity, hence we exclude these solutions (if any) 
	from the reconstruction process, \\
	\item and finally, we generate many reconstructed comoving distances as well as its first and second derivatives, and then compute the cosmokinetic parameters [e.g $H(z)$,$ q(z)$ and $j(z)$]. 
\end{enumerate}


\section{The estimated cosmokinetic parameters}\label{sec:res}
\begin{figure*}
	\centering
	\includegraphics[width=18. cm]{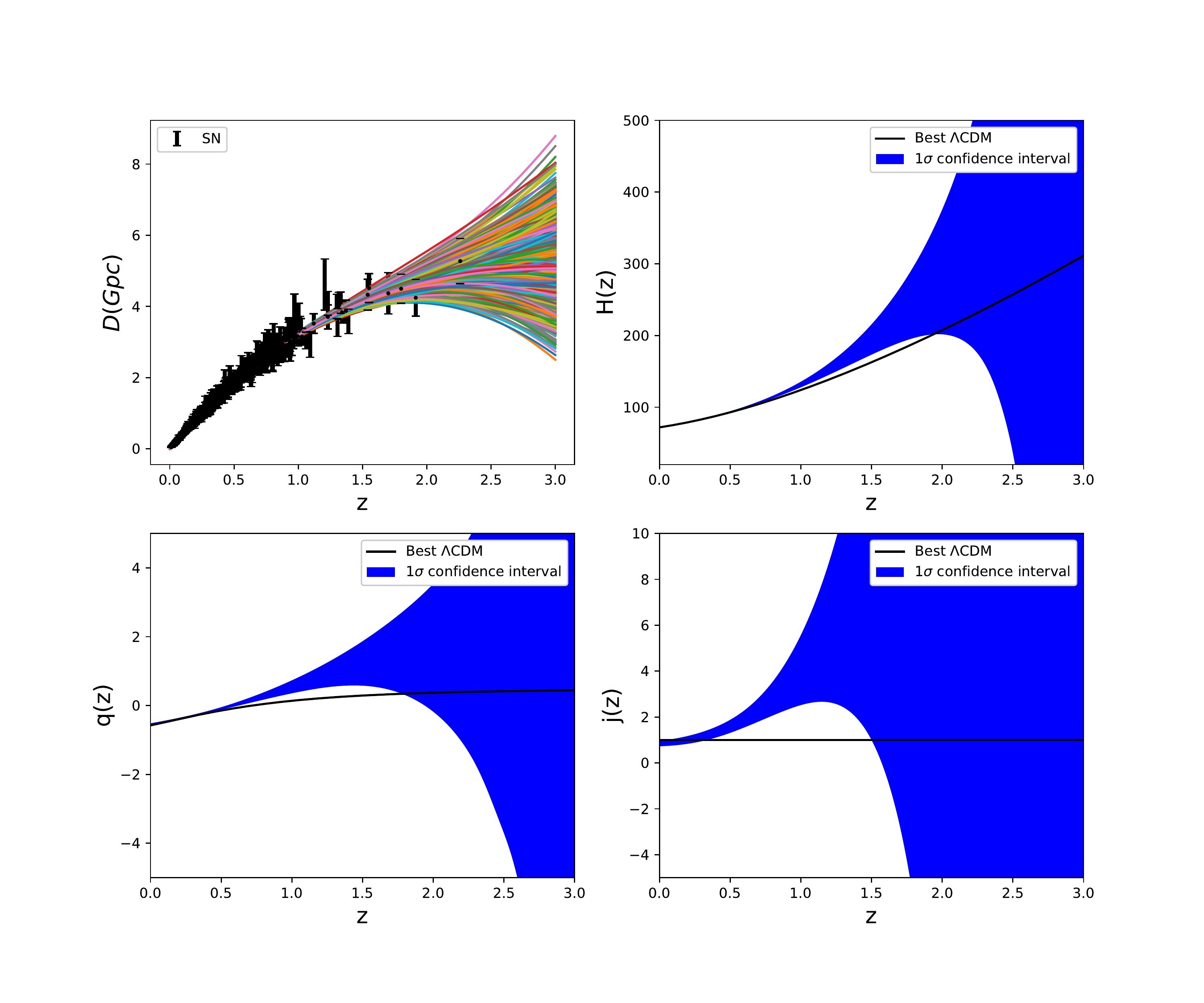}
	\caption{The reconstruction of the 
		cosmokinetic parameters using SNIa data with Gaussian kernel.}
	\label{fig:gu_sn}
\end{figure*}

\begin{figure*}
	\centering
	\includegraphics[width=18 cm]{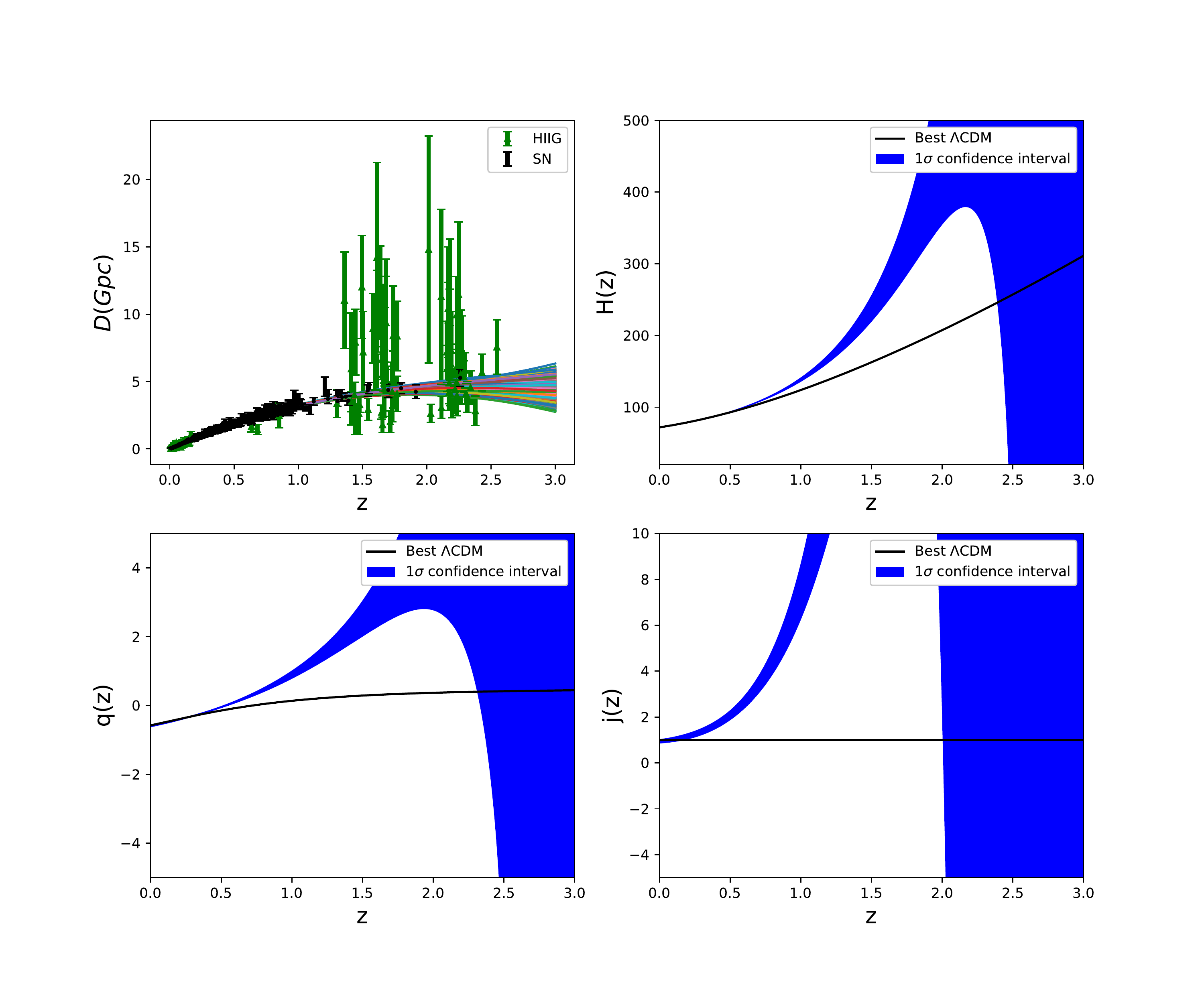}
	\caption{The reconstruction of cosmokinetic parameters using SNIa and HII data with Gaussian kernel.}
	\label{fig:gu_sn_g}
\end{figure*}

In this section we discuss the main results of our analysis,
utilizing the aforementioned data sets and methods. 
In order to reconstruct  $H(z)$, $q(z)$ and $j(z)$, we 
use the $D(z)$, $D'(z)$ and $D''(z)$ functions described 
in section \ref{sec:back}. 
In Figures \ref{fig:gu_sn} and \ref{fig:gu_sn_g} 
we present our results considering the Gaussian kernel. 
We have verified that considering other 
kernels the results remain unaltered.
Having obtained the reconstructed 
cosmokinetic parameters 
we compute the corresponding uncertainties at each redshift. 
Since the values of the reconstructed cosmokinetic parameters follow a
normal distribution at each redshift, their uncertainties are given by the standard deviation at those points.
To check the robustness of the results, we sample 
$D(z)$, $D'(z)$ and $D''(z)$ several times and compute the 
cosmokinetic parameters 
from different samples and  verified that the results are stable.  
\begin{figure}
	\centering
	\includegraphics[width=9.5 cm]{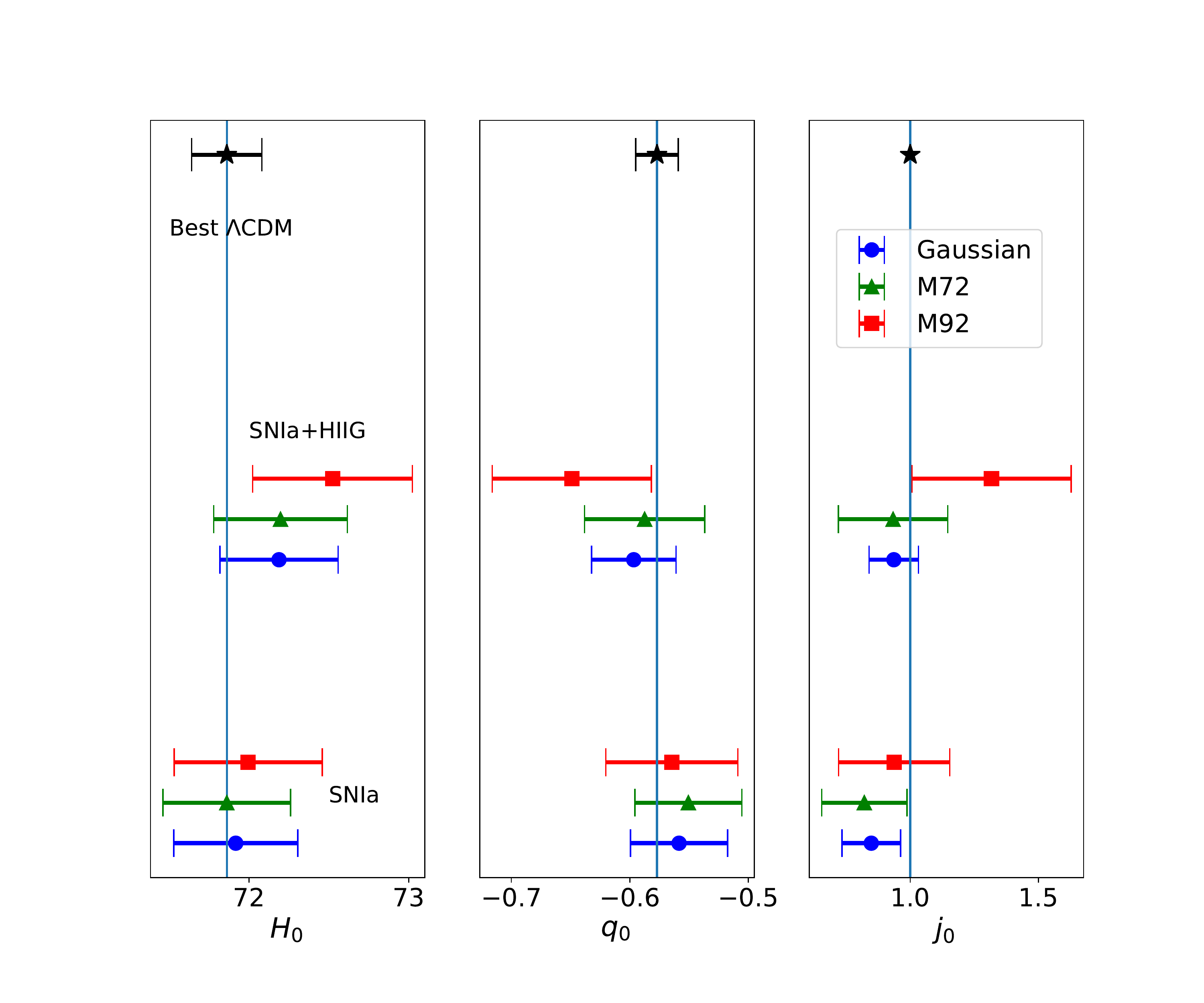}
	\caption{Cosmokinetic parameters at the present time considering different kernels and data sets. Blue circles, green triangles and red squares show the results for Gaussian, Matren $(\nu=3.5)$ and Matren $(\nu=4.5)$, respectively. The best fit $\Lambda$CDM value, considering all the data, is shown by a black star at the top of each panel.}
	\label{fig:H0}
\end{figure} 

In Table (\ref{tab:res}), we provide a summary of the
obtained results. We also
show the extracted cosmokinetics parameters in Figure (\ref{fig:H0}) at the present time.

\begin{table*}
	\begin{center}
		\begin{tabular}{ |c|c|c|c| } 
			\hline
			\diagbox{Data set}{Kernel}  & Gaussian  &  Matren $\nu=7/2$&  Matren $\nu=9/2$ \\
			\hline
			SNIa & \makecell{$H_0$= $71.92\pm 0.38$ $\vspace{0.2 cm}$\\ $q_0$=$-0.558\pm 0.040$ $\vspace{0.2 cm}$ \\ $j_0$=$0.840\pm 0.110$} &  \makecell{$H_0$=$71.86\pm 0.40$ $\vspace{0.2 cm}$ \\ $q_0$=$-0.550\pm 0.045$ $\vspace{0.2 cm}$\\ $j_0$=$0.820\pm 0.170$}& \makecell{$H_0$=$71.99\pm 0.46$ $\vspace{0.2 cm}$ \\ $q_0$=$-0.564\pm 0.050$ $\vspace{0.2 cm}$\\ $j_0$=$0.940\pm 0.220$} \\
			\hline
			SNIa+HIIG & \makecell{$H_0$=$72.18\pm 0.37$  $\vspace{0.2 cm}$\\ $q_0$=$-0.596\pm 0.035$ $\vspace{0.2 cm}$\\ $j_0$=$0.936\pm 0.096$} &  \makecell{$H_0$=$72.19\pm 0.42$ $\vspace{0.2 cm}$ \\ $q_0$=$-0.587\pm 0.050$ $\vspace{0.2 cm}$\\ $j_0$=$0.933\pm 0.213$}
			& \makecell{$H_0$=$72.52 \pm 0.50$ $\vspace{0.2 cm}$ \\ $q_0$=$-0.648\pm 0.067$ $\vspace{0.2 cm}$\\ $j_0$=$1.310\pm0.310$} \\
			\hline
		\end{tabular}
		\caption{Cosmokinetic parameters at the present time for different data sets and kernels. }\label{tab:res}
	\end{center} 
\end{table*}

Overall, we find that our results are in very good agreement with those of 
\cite{Mehrabi_2020} who used Gamma ray bursts instead of HIIG 
and a similar approach to the one presented here.
The current GP analysis shows that close to the present time
we acquire $\sim 1\sigma$ compatibility in all cases. 
Specifically, the dimensionless parameters $q_{0}$ and $j_{0}$  
are constrained to an interval which includes the concordance 
$\Lambda$CDM model and there is only a small deviation 
in $H_0$ considering all data sets.  


\begin{table}
	\begin{center}
		\begin{tabular}{ |c|c|c|c|c| } 
			\hline
			\diagbox{Data set}{Parameter}& Redshifts  & H(z)  &  q(z) &  j(z) \\
			\hline
			SNIa&\makecell{$z=0.5$\\ $z=1$ \\ $z=1.5$} &
			\makecell{~~$1.5$\\ ~~$2.0$ \\ ~~$1.5$} &
			\makecell{$1.6$\\ $2.1$ \\ $1.5$}& 
			\makecell{$2.1$\\ $2.0$ \\ $1.0$}\\ 
			\hline
			SNIa+HIIG&\makecell{$z=0.5$\\ $z=1$ \\ $z=1.5$} &
			\makecell{$1.6$\\ $4.8$ \\ $4.0$} &
			\makecell{$3.2$\\ $5.8$ \\ $4.1$}&
			\makecell{$5.0$\\ $4.9$ \\ $2.5$}\\
			\hline
			
		\end{tabular}
	\end{center}
	\caption{Amount of $\sigma$ deviation ($\Delta X(z)$) from the best $\Lambda$CDM model at 3 different redshifts.}\label{tab:dev} 
\end{table}

Next we focus on the evolution of the cosmokinetic parameters.
For the given cosmological quantity $X(z)$, namely
$H(z)$, $q(z)$, and $j(z)$ we estimate the corresponding deviation $\Delta X$, with respect to the $\Lambda$CDM
solution 
\begin{equation}
\Delta X(z)=\left [ \frac{X(z)-X_{\Lambda}(z)}{\sigma_{X}(z)} \right],
\end{equation}
where $\sigma_{X}(z)$ is the uncertainty of the reconstructed 
$X$ parameter at redshift $z$. 
In Table (\ref{tab:dev}) we provide an overall presentation of the 
relative deviation for three different cosmic epochs, namely $z=0.5,1$ and 1.5. 
Also, in Figures \ref{fig:gu_sn} and \ref{fig:gu_sn_g} 
we show the evolution of the reconstructed cosmokinetic parameters. 
At this point we need to mention 
that for $z>1.5$ the deviation starts to decay due to 
large uncertainties in the reconstruction of the Hubble 
relation and its derivatives, hence we restrict our analysis to the aforementioned cosmic epochs.

In the case of SNIa data, the deviation 
of the reconstructed cosmokinetic
parameters with respect to those of the $\Lambda$CDM lies in the 
interval $[1\sigma,2.1\sigma]$, implying that these data alone do 
not indicate any significant difference. 
These results remain robust regardless of the form of the 
corresponding kernel used during the reconstruction process.
First, including HIIG data, the deviation in all $H(z)$, $q(z)$ and $j(z)$ 
is significant for $z\sim 1$ (or $1.5$).
Specifically, 
close to 
$z=1$ (or $z=1.5$), we find that 
the reconstructed Hubble parameter deviates from the best $H_{\Lambda}(z)$ 
by $\sim 4.8 \sigma, (\sim 4\sigma)$.
Moreover for the total data set, SNIa+HIIG,  
we see that prior to the present epoch, since SNIa data dominate the Hubble relation, 
the reconstructed Hubble parameter tends to that of 
$\Lambda$CDM, namely the relative deviation is quite small $\sim 1.6\sigma$.  
Regarding the deceleration parameter, we observe a clear deviation 
from the $\Lambda$CDM model, which can reach up to 
$\sim 5.8\sigma$ (or $4.1\sigma$) at $z=1$ (or $z=1.5$) 
[see SNIa+HIIG in Table(\ref{tab:dev})].    
Concerning the jerk parameter, there is a visible 
deviation from the $\Lambda$CDM model, where the maximum 
tension is  $\sim 5\sigma$ at $z=0.5$ for the set 
SNIa+HIIG. 

\begin{figure*}
	\centering
	\includegraphics[width=18 cm]{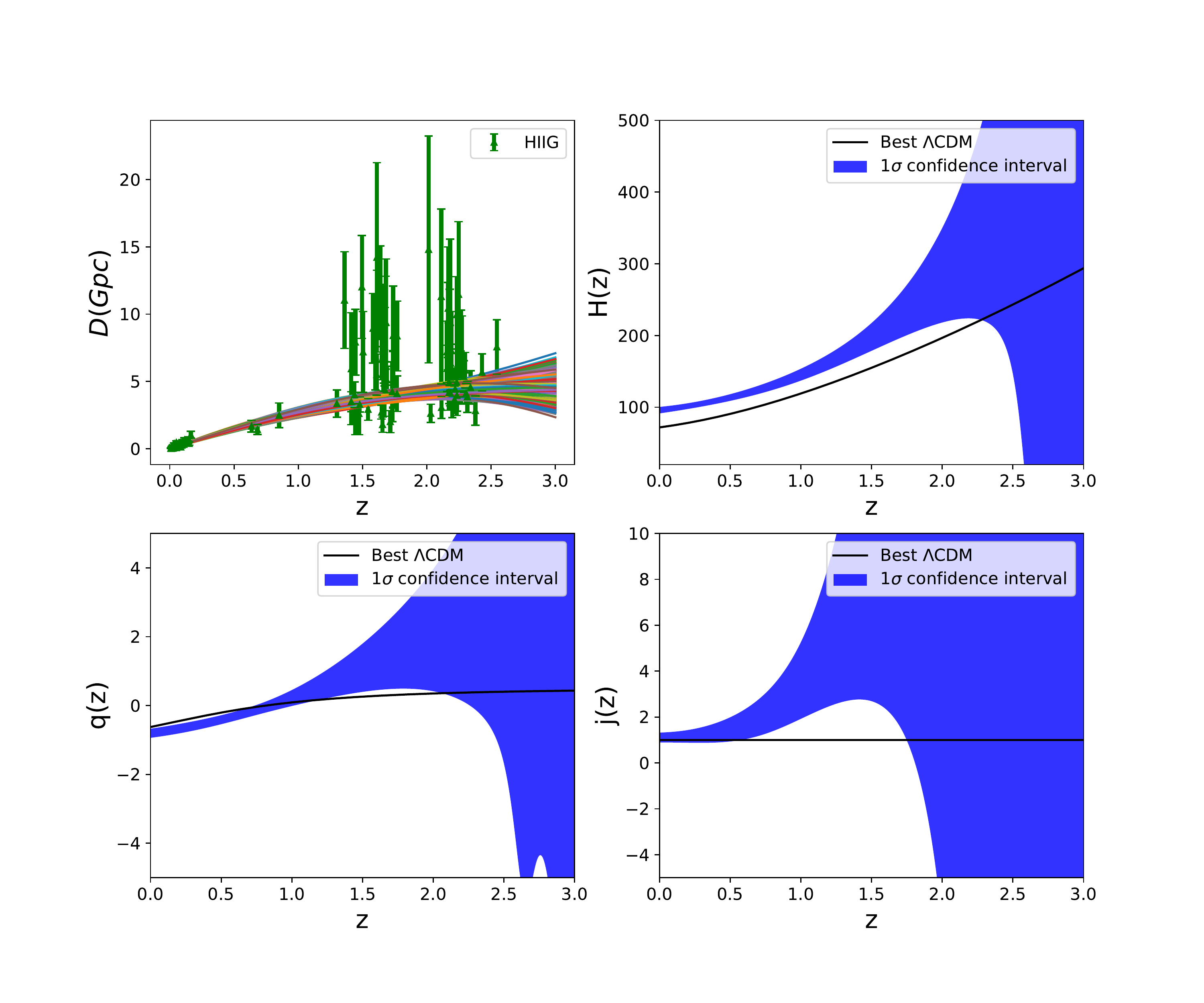}
	\caption{The reconstruction of cosmokinetic parameters using only HIIG data with Gaussian kernel.}
	\label{fig:gu_hg}
\end{figure*}

Finally, we run our code only for HIIG data and in 
Figure (\ref{fig:gu_hg})
we show the corresponding results.
We find that the performance of the 
reconstruction method is rather poor with respect to that for 
SNIa, but this is not surprising because in our case
we have only 181 HIIG, most of them at $z<0.16$ (106 objects)
a region of space where differences between cosmological models 
are almost negligible. 
However, the strength of our results is that the HIIG sample includes 69
objects with $z>1.4$ vs. only 6 SNIa.   
Indeed from Figure (\ref{fig:gu_hg}) we observe that despite the poor reconstruction 
of the cosmic expansion it is encouraging that the jerk 
parameter starts to deviate from unity at 
relatively high-z. This is an 
indication for deviations in the evolution of $j(z)$ with
respect to the expectations of the usual $\Lambda$CDM model.
Of course as we have already discussed above
combining HIIG with SNIa we obtain similar results to 
those of \cite{Lusso_2020,Mehrabi_2020}, where 
they have combined SNIa with 
other potential high-z cosmic tracers (quasars, GRBs)
in order to extend the Hubble relation to as high redshifts as possible.

\section{Discussion and Conclusions}\label{sec:con}
Although the concordance $\Lambda$CDM model is considered 
to be in agreement with the majority of 
observational data, lately there have been heated debates
in the literature since it seems to be in tension with some recent measurements,
related to the Hubble constant $H_{0}$ and the current
value of the mass variance in spheres of 8 $h^{-1}$ Mpc radius \citep{Sola:2017znb,Verde_2019,DIVALENTINO2021102604,DIVALENTINO2021102605,Perivolaropoulos:2021jda}.

From the view point of the Hubble diagram, 
\cite{Lusso:2019akb,Lusso_2020} combining 
the nominal standard candles (SNIa) with other probes such as
quasars and Gamma-Ray Bursts obtained a $\sim 4\sigma$ tension between the 
best fit cosmokinetic parameters and the $\Lambda$CDM expectations
(see also \citet{Lusso_2020,Mehrabi_2020}).
Whether these tensions are the result of yet unknown systematic
errors or hint towards new Physics is still unclear. 
Therefore, within the framework of the traditional $\Lambda$CDM model, 
testing the validity of the Hubble diagram at large redshifts 
is considered one of the most crucial tasks in cosmology, hence it is 
important to minimize the amount of assumptions 
needed to successfully complete such an effort. 
One possible avenue is to reconstruct the Hubble diagram  
in a model independent way, and thence to estimate the cosmokinetic parameters. 

In this article we worked along the above lines 
and in particular we used HIIG as alternative 
tracers of the Hubble expansion and combined them 
with the SNIa 
data aiming to reconstruct the Hubble diagram in a model independent way 
and thus to constrain the corresponding
cosmokinetic parameters. We performed a reconstruction method 
which is based on the so called GP 
method \citep{GP_book,Seikel:2012uu,Mehrabi_2020}
in the context of three different kernels. 
In the case of SNIa data 
we found that the cosmokinetic parameters, extracted from the  
Gaussian process, are consistent with those of $\Lambda$CDM at 
the present time. 
Including HIIG 
in the analysis, 
we didn't find significant deviations 
from $\Lambda$CDM, as far as the current value of 
cosmokinetic parameters is concerned. 

Then we focused on the  evolution of the cosmokinetic parameters 
and we verified that it is not significantly affected 
by the choice of the kernel. 
Since our method has some limitations due to large uncertainties in
reconstructing properly the Hubble relation at large enough redshits, we have 
restricted our analysis to intermediate cosmic epochs, namely $z\in [1, 1.5]$. 
For this redshift interval we found that the estimated cosmokinetic parameters
significantly deviate from those of the usual $\Lambda$CDM model.
Indeed 
for the SNIa+HIIG 
combination, the 
corresponding deviation, close to $z=1.5$, can reach up 
to $\sim 4 \sigma$ in the case of $H(z)$, $\sim 4.1\sigma$ in the case of 
$q(z)$ and $\sim 2.5\sigma$ for $j(z)$, while for $z=1$ the corresponding 
deviations become even larger.

In a nutshell, we argue that the potential benefit of having 
standard tracers for
cosmology at high redshifts makes it tempting 
to try and use HIIG to 
constrain the cosmological expansion history in a way similar to
SNIa. Notice that our team has thoroughly studied 
in a series of 
papers (see \cite{Plionis:2011jj,Terlevich:2015toa})
the question of which is the most efficient strategy 
to tighten the cosmological constraints provided by fitting 
the Hubble relation. 
Using extensive Monte-Carlo simulations we have found 
that by using a number of $\sim 300$ high z tracers, even with a
relatively large distance modulus uncertainty, we can 
reduce significantly the cosmological parameter solution space.
Currently, our group 
is designing the appropriate KMOS-VLT observations of 
high-z objects aiming to include 
$\sim$100 additional entries in the HIIG sample 
which are expected to substantially 
improve  the  performance  of  our  method  in  reconstructing  (in
a model-independent way) the Hubble relation, hence to 
use cosmokinetics in order to test the validity of the concordance $\Lambda$CDM.

The use of HIIG as distance indicators and to deduce 
cosmological parameters ($\Omega_{m0}$, $H_{0}$, equation of state 
parameter etc) is not new. 
Since the work, e.g., of \cite{10.1093/mnras/235.1.297}, our group has 
refined the method and, following e.g. \cite{Plionis:2011jj,Terlevich:2015toa,Chavez:2016epc} simulations, have embarked in a 
concerted observational programme destined to improve 
the pool of data, in particular at higher z ranges, where differences 
between cosmological models are more prominent, and on better understanding 
the systematic errors in the method. 
Specifically, \cite{Wei:2016jqa} have used the data to investigate 
performance of some well-known models including 
the $\Lambda$CDM, wCDM (quintessence) and $R_h=ct$ universe. Their analysis 
indicated that the $R_h=ct$ model is strongly favored 
over wCDM. 
On the other hand, \cite{Yennapureddy:2017vvb} have 
used a sub-sample of our HIIG data and the GP method 
in order to reconstruct the Hubble relation, without estimating 
the cosmokinetic parameters.
In addition, in a more recent work \cite{Cao:2021cix}, 
have combined two compilations of HIIG data with quasars, BAOs, cosmic expansion $H(z)$ data 
and SNIa data in order to place constrains on several cosmological models.


Overall, we argue that by including high-$z$ alternative tracers in the Hubble
diagram, there is a discrepancy/tension between the measured cosmokinetic parameters
with those of the concordance $\Lambda$CDM model in intermediate cosmic epochs,
possibly an indication for new Physics, if one would eventually exclude
systematics and small-number statistics.

Therefore, in order to clarify the situation it is necessary 
to increase our HIIG sample, especially at high redshifts. 
Indeed this has been shown to be crucial, since 
according to \citep{Plionis:2011jj,Chavez:2016epc}, a 
possible deviation
from $\Lambda$CDM could in principle be detected when using a
few hundreds of high redshift HIIG in the $1.5 \le z \le 4$ interval.

In the current work we used the newest HIIG sample that contains in total 181
entries, out of which 74 are high redshift HIIG, already providing very
promising results. In the future our group is designing the appropriate 
KMOS-VLT observations of high-z objects aiming to add $\sim$ 100
additional HIIG in the sample which are expected to substantially
improve the performance of our method in reconstructing (in a 
model-independent way) the Hubble relation, and thus 
the validity of the concordance $\Lambda$CDM model will
be effectively tested via the cosmokinetic analysis.

   



 \bibliographystyle{apsrev4-1}
\bibliography{ref.bib}

\begin{thebibliography}{61}%
\makeatletter
\providecommand \@ifxundefined [1]{%
 \@ifx{#1\undefined}
}%
\providecommand \@ifnum [1]{%
 \ifnum #1\expandafter \@firstoftwo
 \else \expandafter \@secondoftwo
 \fi
}%
\providecommand \@ifx [1]{%
 \ifx #1\expandafter \@firstoftwo
 \else \expandafter \@secondoftwo
 \fi
}%
\providecommand \natexlab [1]{#1}%
\providecommand \enquote  [1]{``#1''}%
\providecommand \bibnamefont  [1]{#1}%
\providecommand \bibfnamefont [1]{#1}%
\providecommand \citenamefont [1]{#1}%
\providecommand \href@noop [0]{\@secondoftwo}%
\providecommand \href [0]{\begingroup \@sanitize@url \@href}%
\providecommand \@href[1]{\@@startlink{#1}\@@href}%
\providecommand \@@href[1]{\endgroup#1\@@endlink}%
\providecommand \@sanitize@url [0]{\catcode `\\12\catcode `\$12\catcode
  `\&12\catcode `\#12\catcode `\^12\catcode `\_12\catcode `\%12\relax}%
\providecommand \@@startlink[1]{}%
\providecommand \@@endlink[0]{}%
\providecommand \url  [0]{\begingroup\@sanitize@url \@url }%
\providecommand \@url [1]{\endgroup\@href {#1}{\urlprefix }}%
\providecommand \urlprefix  [0]{URL }%
\providecommand \Eprint [0]{\href }%
\providecommand \doibase [0]{http://dx.doi.org/}%
\providecommand \selectlanguage [0]{\@gobble}%
\providecommand \bibinfo  [0]{\@secondoftwo}%
\providecommand \bibfield  [0]{\@secondoftwo}%
\providecommand \translation [1]{[#1]}%
\providecommand \BibitemOpen [0]{}%
\providecommand \bibitemStop [0]{}%
\providecommand \bibitemNoStop [0]{.\EOS\space}%
\providecommand \EOS [0]{\spacefactor3000\relax}%
\providecommand \BibitemShut  [1]{\csname bibitem#1\endcsname}%
\let\auto@bib@innerbib\@empty
\bibitem [{\citenamefont {Riess}\ \emph {et~al.}(1998)\citenamefont {Riess},
  \citenamefont {Filippenko}, \citenamefont {Challis},\ and\ \citenamefont
  {et~al.}}]{Riess1998}%
  \BibitemOpen
  \bibfield  {author} {\bibinfo {author} {\bibfnamefont {A.~G.}\ \bibnamefont
  {Riess}}, \bibinfo {author} {\bibfnamefont {A.~V.}\ \bibnamefont
  {Filippenko}}, \bibinfo {author} {\bibfnamefont {P.}~\bibnamefont {Challis}},
  \ and\ \bibinfo {author} {\bibnamefont {et~al.}},\ }\href@noop {} {\bibfield
  {journal} {\bibinfo  {journal} {\aj}\ }\textbf {\bibinfo {volume} {116}},\
  \bibinfo {pages} {1009} (\bibinfo {year} {1998})}\BibitemShut {NoStop}%
\bibitem [{\citenamefont {Perlmutter}\ \emph {et~al.}(1999)\citenamefont
  {Perlmutter}, \citenamefont {Aldering}, \citenamefont {Goldhaber},\ and\
  \citenamefont {et~al.}}]{Perlmutter1999}%
  \BibitemOpen
  \bibfield  {author} {\bibinfo {author} {\bibfnamefont {S.}~\bibnamefont
  {Perlmutter}}, \bibinfo {author} {\bibfnamefont {G.}~\bibnamefont
  {Aldering}}, \bibinfo {author} {\bibfnamefont {G.}~\bibnamefont {Goldhaber}},
  \ and\ \bibinfo {author} {\bibnamefont {et~al.}},\ }\href@noop {} {\bibfield
  {journal} {\bibinfo  {journal} {\apj}\ }\textbf {\bibinfo {volume} {517}},\
  \bibinfo {pages} {565} (\bibinfo {year} {1999})}\BibitemShut {NoStop}%
\bibitem [{\citenamefont {Komatsu}\ \emph {et~al.}(2011)\citenamefont
  {Komatsu}, \citenamefont {Smith}, \citenamefont {Dunkley},\ and\
  \citenamefont {et~al.}}]{Komatsu2011}%
  \BibitemOpen
  \bibfield  {author} {\bibinfo {author} {\bibfnamefont {E.}~\bibnamefont
  {Komatsu}}, \bibinfo {author} {\bibfnamefont {K.~M.}\ \bibnamefont {Smith}},
  \bibinfo {author} {\bibfnamefont {J.}~\bibnamefont {Dunkley}}, \ and\
  \bibinfo {author} {\bibnamefont {et~al.}},\ }\href@noop {} {\bibfield
  {journal} {\bibinfo  {journal} {\apjs}\ }\textbf {\bibinfo {volume} {192}},\
  \bibinfo {pages} {18} (\bibinfo {year} {2011})}\BibitemShut {NoStop}%
\bibitem [{\citenamefont {{Planck Collaboration XIV}}(2016)}]{Ade:2015yua}%
  \BibitemOpen
  \bibfield  {author} {\bibinfo {author} {\bibnamefont {{Planck Collaboration
  XIV}}} (\bibinfo {collaboration} {Planck Collaboration}),\ }\href@noop {}
  {\bibfield  {journal} {\bibinfo  {journal} {Astron.Astrophys.}\ }\textbf
  {\bibinfo {volume} {594}},\ \bibinfo {pages} {A14} (\bibinfo {year}
  {2016})}\BibitemShut {NoStop}%
\bibitem [{\citenamefont {Aghanim}\ \emph {et~al.}(2018)\citenamefont {Aghanim}
  \emph {et~al.}}]{Aghanim:2018eyx}%
  \BibitemOpen
  \bibfield  {author} {\bibinfo {author} {\bibfnamefont {N.}~\bibnamefont
  {Aghanim}} \emph {et~al.} (\bibinfo {collaboration} {Planck}),\ }\href@noop
  {} {\  (\bibinfo {year} {2018})},\ \Eprint {http://arxiv.org/abs/1807.06209}
  {arXiv:1807.06209 [astro-ph.CO]} \BibitemShut {NoStop}%
\bibitem [{\citenamefont {Eisenstein}\ \emph {et~al.}(2005)\citenamefont
  {Eisenstein} \emph {et~al.}}]{Eisenstein:2005su}%
  \BibitemOpen
  \bibfield  {author} {\bibinfo {author} {\bibfnamefont {D.~J.}\ \bibnamefont
  {Eisenstein}} \emph {et~al.} (\bibinfo {collaboration} {SDSS
  Collaboration}),\ }\href {\doibase 10.1086/466512} {\bibfield  {journal}
  {\bibinfo  {journal} {ApJ}\ }\textbf {\bibinfo {volume} {633}},\ \bibinfo
  {pages} {560} (\bibinfo {year} {2005})}\BibitemShut {NoStop}%
\bibitem [{\citenamefont {Percival}\ \emph {et~al.}(2010)\citenamefont
  {Percival}, \citenamefont {Reid}, \citenamefont {Eisenstein},\ and\
  \citenamefont {et~al.}}]{Percival2010}%
  \BibitemOpen
  \bibfield  {author} {\bibinfo {author} {\bibfnamefont {W.~J.}\ \bibnamefont
  {Percival}}, \bibinfo {author} {\bibfnamefont {B.~A.}\ \bibnamefont {Reid}},
  \bibinfo {author} {\bibfnamefont {D.~J.}\ \bibnamefont {Eisenstein}}, \ and\
  \bibinfo {author} {\bibnamefont {et~al.}},\ }\href@noop {} {\bibfield
  {journal} {\bibinfo  {journal} {\mnras}\ }\textbf {\bibinfo {volume} {401}},\
  \bibinfo {pages} {2148} (\bibinfo {year} {2010})}\BibitemShut {NoStop}%
\bibitem [{\citenamefont {Blake}\ \emph {et~al.}(2011)\citenamefont {Blake}
  \emph {et~al.}}]{Blake:2011rj}%
  \BibitemOpen
  \bibfield  {author} {\bibinfo {author} {\bibfnamefont {C.}~\bibnamefont
  {Blake}} \emph {et~al.},\ }\href {\doibase 10.1111/j.1365-2966.2011.18903.x}
  {\bibfield  {journal} {\bibinfo  {journal} {Mon. Not. Roy. Astron. Soc.}\
  }\textbf {\bibinfo {volume} {415}},\ \bibinfo {pages} {2876} (\bibinfo {year}
  {2011})},\ \Eprint {http://arxiv.org/abs/1104.2948} {arXiv:1104.2948
  [astro-ph.CO]} \BibitemShut {NoStop}%
\bibitem [{\citenamefont {Reid}\ \emph {et~al.}(2012)\citenamefont {Reid},
  \citenamefont {Samushia}, \citenamefont {White}, \citenamefont {Percival},
  \citenamefont {Manera} \emph {et~al.}}]{Reid:2012sw}%
  \BibitemOpen
  \bibfield  {author} {\bibinfo {author} {\bibfnamefont {B.~A.}\ \bibnamefont
  {Reid}}, \bibinfo {author} {\bibfnamefont {L.}~\bibnamefont {Samushia}},
  \bibinfo {author} {\bibfnamefont {M.}~\bibnamefont {White}}, \bibinfo
  {author} {\bibfnamefont {W.~J.}\ \bibnamefont {Percival}}, \bibinfo {author}
  {\bibfnamefont {M.}~\bibnamefont {Manera}},  \emph {et~al.},\ }\href
  {\doibase 10.1111/j.1365-2966.2012.21779.x} {\bibfield  {journal} {\bibinfo
  {journal} {MNRAS}\ }\textbf {\bibinfo {volume} {426}},\ \bibinfo {pages}
  {2719} (\bibinfo {year} {2012})}\BibitemShut {NoStop}%
\bibitem [{\citenamefont {Abbott}\ \emph {et~al.}(2019)\citenamefont {Abbott}
  \emph {et~al.}}]{Abbott:2017wcz}%
  \BibitemOpen
  \bibfield  {author} {\bibinfo {author} {\bibfnamefont {T.~M.~C.}\
  \bibnamefont {Abbott}} \emph {et~al.} (\bibinfo {collaboration} {DES}),\
  }\href {\doibase 10.1093/mnras/sty3351} {\bibfield  {journal} {\bibinfo
  {journal} {Mon. Not. Roy. Astron. Soc.}\ }\textbf {\bibinfo {volume} {483}},\
  \bibinfo {pages} {4866} (\bibinfo {year} {2019})},\ \Eprint
  {http://arxiv.org/abs/1712.06209} {arXiv:1712.06209 [astro-ph.CO]}
  \BibitemShut {NoStop}%
\bibitem [{\citenamefont {Alam}\ \emph {et~al.}(2017)\citenamefont {Alam} \emph
  {et~al.}}]{Alam:2016hwk}%
  \BibitemOpen
  \bibfield  {author} {\bibinfo {author} {\bibfnamefont {S.}~\bibnamefont
  {Alam}} \emph {et~al.} (\bibinfo {collaboration} {BOSS}),\ }\href {\doibase
  10.1093/mnras/stx721} {\bibfield  {journal} {\bibinfo  {journal} {Mon. Not.
  Roy. Astron. Soc.}\ }\textbf {\bibinfo {volume} {470}},\ \bibinfo {pages}
  {2617} (\bibinfo {year} {2017})},\ \Eprint {http://arxiv.org/abs/1607.03155}
  {arXiv:1607.03155 [astro-ph.CO]} \BibitemShut {NoStop}%
\bibitem [{\citenamefont {Gil-Marín}\ \emph {et~al.}(2018)\citenamefont
  {Gil-Marín} \emph {et~al.}}]{Gil-Marin:2018cgo}%
  \BibitemOpen
  \bibfield  {author} {\bibinfo {author} {\bibfnamefont {H.}~\bibnamefont
  {Gil-Marín}} \emph {et~al.},\ }\href {\doibase 10.1093/mnras/sty453}
  {\bibfield  {journal} {\bibinfo  {journal} {Mon. Not. Roy. Astron. Soc.}\
  }\textbf {\bibinfo {volume} {477}},\ \bibinfo {pages} {1604} (\bibinfo {year}
  {2018})},\ \Eprint {http://arxiv.org/abs/1801.02689} {arXiv:1801.02689
  [astro-ph.CO]} \BibitemShut {NoStop}%
\bibitem [{\citenamefont {Farooq}\ \emph {et~al.}(2017)\citenamefont {Farooq},
  \citenamefont {Madiyar}, \citenamefont {Crandall},\ and\ \citenamefont
  {Ratra}}]{Farooq:2016zwm}%
  \BibitemOpen
  \bibfield  {author} {\bibinfo {author} {\bibfnamefont {O.}~\bibnamefont
  {Farooq}}, \bibinfo {author} {\bibfnamefont {F.~R.}\ \bibnamefont {Madiyar}},
  \bibinfo {author} {\bibfnamefont {S.}~\bibnamefont {Crandall}}, \ and\
  \bibinfo {author} {\bibfnamefont {B.}~\bibnamefont {Ratra}},\ }\href
  {\doibase 10.3847/1538-4357/835/1/26} {\bibfield  {journal} {\bibinfo
  {journal} {Astrophys. J.}\ }\textbf {\bibinfo {volume} {835}},\ \bibinfo
  {pages} {26} (\bibinfo {year} {2017})},\ \Eprint
  {http://arxiv.org/abs/1607.03537} {arXiv:1607.03537 [astro-ph.CO]}
  \BibitemShut {NoStop}%
\bibitem [{\citenamefont {Weinberg}(1989)}]{Weinberg:1989}%
  \BibitemOpen
  \bibfield  {author} {\bibinfo {author} {\bibfnamefont {S.}~\bibnamefont
  {Weinberg}},\ }\href {\doibase 10.1103/RevModPhys.61.1} {\bibfield  {journal}
  {\bibinfo  {journal} {Reviews of Modern Physics}\ }\textbf {\bibinfo {volume}
  {61}},\ \bibinfo {pages} {1} (\bibinfo {year} {1989})}\BibitemShut {NoStop}%
\bibitem [{\citenamefont {Peebles}\ and\ \citenamefont
  {Ratra}(2003)}]{Peebles:2002gy}%
  \BibitemOpen
  \bibfield  {author} {\bibinfo {author} {\bibfnamefont {P.}~\bibnamefont
  {Peebles}}\ and\ \bibinfo {author} {\bibfnamefont {B.}~\bibnamefont
  {Ratra}},\ }\href {\doibase 10.1103/RevModPhys.75.559} {\bibfield  {journal}
  {\bibinfo  {journal} {Rev. Mod. Phys.}\ }\textbf {\bibinfo {volume} {75}},\
  \bibinfo {pages} {559} (\bibinfo {year} {2003})}\BibitemShut {NoStop}%
\bibitem [{\citenamefont {Copeland}\ \emph {et~al.}(2006)\citenamefont
  {Copeland}, \citenamefont {Sami},\ and\ \citenamefont
  {Tsujikawa}}]{Copeland:2006wr}%
  \BibitemOpen
  \bibfield  {author} {\bibinfo {author} {\bibfnamefont {E.~J.}\ \bibnamefont
  {Copeland}}, \bibinfo {author} {\bibfnamefont {M.}~\bibnamefont {Sami}}, \
  and\ \bibinfo {author} {\bibfnamefont {S.}~\bibnamefont {Tsujikawa}},\ }\href
  {\doibase 10.1142/S021827180600942X} {\bibfield  {journal} {\bibinfo
  {journal} {IJMP}\ }\textbf {\bibinfo {volume} {D15}},\ \bibinfo {pages}
  {1753} (\bibinfo {year} {2006})}\BibitemShut {NoStop}%
\bibitem [{\citenamefont {Chiba}\ \emph {et~al.}(2009)\citenamefont {Chiba},
  \citenamefont {Dutta},\ and\ \citenamefont {Scherrer}}]{Chiba:2009nh}%
  \BibitemOpen
  \bibfield  {author} {\bibinfo {author} {\bibfnamefont {T.}~\bibnamefont
  {Chiba}}, \bibinfo {author} {\bibfnamefont {S.}~\bibnamefont {Dutta}}, \ and\
  \bibinfo {author} {\bibfnamefont {R.~J.}\ \bibnamefont {Scherrer}},\ }\href
  {\doibase 10.1103/PhysRevD.80.043517} {\bibfield  {journal} {\bibinfo
  {journal} {Phys. Rev. D}\ }\textbf {\bibinfo {volume} {80}},\ \bibinfo
  {pages} {043517} (\bibinfo {year} {2009})}\BibitemShut {NoStop}%
\bibitem [{\citenamefont {Amendola}\ and\ \citenamefont
  {Tsujikawa}(2010)}]{Amendola:2010}%
  \BibitemOpen
  \bibfield  {author} {\bibinfo {author} {\bibfnamefont {L.}~\bibnamefont
  {Amendola}}\ and\ \bibinfo {author} {\bibfnamefont {S.}~\bibnamefont
  {Tsujikawa}},\ }\href@noop {} {\emph {\bibinfo {title} {Dark Energy: Theory
  and Observations}}}\ (\bibinfo  {publisher} {Cambridge University Press},\
  \bibinfo {address} {Cambridge UK},\ \bibinfo {year} {2010})\BibitemShut
  {NoStop}%
\bibitem [{\citenamefont {Mehrabi}(2018)}]{Mehrabi:2018dru}%
  \BibitemOpen
  \bibfield  {author} {\bibinfo {author} {\bibfnamefont {A.}~\bibnamefont
  {Mehrabi}},\ }\href {\doibase 10.1103/PhysRevD.97.083522} {\bibfield
  {journal} {\bibinfo  {journal} {Phys. Rev.}\ }\textbf {\bibinfo {volume}
  {D97}},\ \bibinfo {pages} {083522} (\bibinfo {year} {2018})},\ \Eprint
  {http://arxiv.org/abs/1804.09886} {arXiv:1804.09886 [astro-ph.CO]}
  \BibitemShut {NoStop}%
\bibitem [{\citenamefont {Mehrabi}\ and\ \citenamefont
  {Basilakos}(2018)}]{Mehrabi:2018oke}%
  \BibitemOpen
  \bibfield  {author} {\bibinfo {author} {\bibfnamefont {A.}~\bibnamefont
  {Mehrabi}}\ and\ \bibinfo {author} {\bibfnamefont {S.}~\bibnamefont
  {Basilakos}},\ }\href {\doibase 10.1140/epjc/s10052-018-6368-x} {\bibfield
  {journal} {\bibinfo  {journal} {Eur. Phys. J.}\ }\textbf {\bibinfo {volume}
  {C78}},\ \bibinfo {pages} {889} (\bibinfo {year} {2018})},\ \Eprint
  {http://arxiv.org/abs/1804.10794} {arXiv:1804.10794 [astro-ph.CO]}
  \BibitemShut {NoStop}%
\bibitem [{\citenamefont {Schmidt}(1990)}]{Schmidt:1990gb}%
  \BibitemOpen
  \bibfield  {author} {\bibinfo {author} {\bibfnamefont {H.-J.}\ \bibnamefont
  {Schmidt}},\ }\href@noop {} {\bibfield  {journal} {\bibinfo  {journal}
  {Astron. Nachr.}\ }\textbf {\bibinfo {volume} {311}},\ \bibinfo {pages} {165}
  (\bibinfo {year} {1990})}\BibitemShut {NoStop}%
\bibitem [{\citenamefont {Magnano}\ and\ \citenamefont
  {Sokolowski}(1994)}]{Magnano:1993bd}%
  \BibitemOpen
  \bibfield  {author} {\bibinfo {author} {\bibfnamefont {G.}~\bibnamefont
  {Magnano}}\ and\ \bibinfo {author} {\bibfnamefont {L.~M.}\ \bibnamefont
  {Sokolowski}},\ }\href {\doibase 10.1103/PhysRevD.50.5039} {\bibfield
  {journal} {\bibinfo  {journal} {Phys. Rev. D}\ }\textbf {\bibinfo {volume}
  {50}},\ \bibinfo {pages} {5039} (\bibinfo {year} {1994})}\BibitemShut
  {NoStop}%
\bibitem [{\citenamefont {Dobado}\ and\ \citenamefont
  {Maroto}(1995)}]{Dobado:1994qp}%
  \BibitemOpen
  \bibfield  {author} {\bibinfo {author} {\bibfnamefont {A.}~\bibnamefont
  {Dobado}}\ and\ \bibinfo {author} {\bibfnamefont {A.~L.}\ \bibnamefont
  {Maroto}},\ }\href {\doibase 10.1103/PhysRevD.53.2262,
  10.1103/PhysRevD.52.1895, 10.1103/PhysRevD.52.1895 10.1103/PhysRevD.53.2262}
  {\bibfield  {journal} {\bibinfo  {journal} {Phys. Rev. D}\ }\textbf {\bibinfo
  {volume} {52}},\ \bibinfo {pages} {1895} (\bibinfo {year}
  {1995})}\BibitemShut {NoStop}%
\bibitem [{\citenamefont {Capozziello}\ \emph {et~al.}(2003)\citenamefont
  {Capozziello}, \citenamefont {Carloni},\ and\ \citenamefont
  {Troisi}}]{Capozziello:2003tk}%
  \BibitemOpen
  \bibfield  {author} {\bibinfo {author} {\bibfnamefont {S.}~\bibnamefont
  {Capozziello}}, \bibinfo {author} {\bibfnamefont {S.}~\bibnamefont
  {Carloni}}, \ and\ \bibinfo {author} {\bibfnamefont {A.}~\bibnamefont
  {Troisi}},\ }\href@noop {} {\bibfield  {journal} {\bibinfo  {journal} {Recent
  Res. Dev. Astron. Astrophys.}\ }\textbf {\bibinfo {volume} {1}},\ \bibinfo
  {pages} {625} (\bibinfo {year} {2003})}\BibitemShut {NoStop}%
\bibitem [{\citenamefont {Carroll}\ \emph {et~al.}(2004)\citenamefont
  {Carroll}, \citenamefont {Duvvuri}, \citenamefont {Trodden},\ and\
  \citenamefont {Turner}}]{Carroll:2003wy}%
  \BibitemOpen
  \bibfield  {author} {\bibinfo {author} {\bibfnamefont {S.~M.}\ \bibnamefont
  {Carroll}}, \bibinfo {author} {\bibfnamefont {V.}~\bibnamefont {Duvvuri}},
  \bibinfo {author} {\bibfnamefont {M.}~\bibnamefont {Trodden}}, \ and\
  \bibinfo {author} {\bibfnamefont {M.~S.}\ \bibnamefont {Turner}},\ }\href
  {\doibase 10.1103/PhysRevD.70.043528} {\bibfield  {journal} {\bibinfo
  {journal} {Phys. Rev. D}\ }\textbf {\bibinfo {volume} {70}},\ \bibinfo
  {pages} {043528} (\bibinfo {year} {2004})}\BibitemShut {NoStop}%
\bibitem [{\citenamefont {{Weinberg}}(1989)}]{1989RvMP...61....1W}%
  \BibitemOpen
  \bibfield  {author} {\bibinfo {author} {\bibfnamefont {S.}~\bibnamefont
  {{Weinberg}}},\ }\href {\doibase 10.1103/RevModPhys.61.1} {\bibfield
  {journal} {\bibinfo  {journal} {Reviews of Modern Physics}\ }\textbf
  {\bibinfo {volume} {61}},\ \bibinfo {pages} {1} (\bibinfo {year}
  {1989})}\BibitemShut {NoStop}%
\bibitem [{\citenamefont {{Padmanabhan}}(2003)}]{2003PhR...380..235P}%
  \BibitemOpen
  \bibfield  {author} {\bibinfo {author} {\bibfnamefont {T.}~\bibnamefont
  {{Padmanabhan}}},\ }\href {\doibase 10.1016/S0370-1573(03)00120-0} {\bibfield
   {journal} {\bibinfo  {journal} {\physrep}\ }\textbf {\bibinfo {volume}
  {380}},\ \bibinfo {pages} {235} (\bibinfo {year} {2003})},\ \Eprint
  {http://arxiv.org/abs/hep-th/0212290} {arXiv:hep-th/0212290 [hep-th]}
  \BibitemShut {NoStop}%
\bibitem [{\citenamefont
  {Perivolaropoulos}(2008)}]{perivolaropoulos2008puzzles}%
  \BibitemOpen
  \bibfield  {author} {\bibinfo {author} {\bibfnamefont {L.}~\bibnamefont
  {Perivolaropoulos}},\ }\href@noop {} {\enquote {\bibinfo {title} {Six puzzles
  for lcdm cosmology},}\ } (\bibinfo {year} {2008}),\ \Eprint
  {http://arxiv.org/abs/0811.4684} {arXiv:0811.4684 [astro-ph]} \BibitemShut
  {NoStop}%
\bibitem [{\citenamefont {Padilla}(2015)}]{padilla2015lectures}%
  \BibitemOpen
  \bibfield  {author} {\bibinfo {author} {\bibfnamefont {A.}~\bibnamefont
  {Padilla}},\ }\href@noop {} {\enquote {\bibinfo {title} {Lectures on the
  cosmological constant problem},}\ } (\bibinfo {year} {2015}),\ \Eprint
  {http://arxiv.org/abs/1502.05296} {arXiv:1502.05296 [hep-th]} \BibitemShut
  {NoStop}%
\bibitem [{\citenamefont {Verde}\ \emph {et~al.}(2019)\citenamefont {Verde},
  \citenamefont {Treu},\ and\ \citenamefont {Riess}}]{Verde_2019}%
  \BibitemOpen
  \bibfield  {author} {\bibinfo {author} {\bibfnamefont {L.}~\bibnamefont
  {Verde}}, \bibinfo {author} {\bibfnamefont {T.}~\bibnamefont {Treu}}, \ and\
  \bibinfo {author} {\bibfnamefont {A.~G.}\ \bibnamefont {Riess}},\ }\href
  {\doibase 10.1038/s41550-019-0902-0} {\bibfield  {journal} {\bibinfo
  {journal} {Nature Astronomy}\ }\textbf {\bibinfo {volume} {3}},\ \bibinfo
  {pages} {891–895} (\bibinfo {year} {2019})}\BibitemShut {NoStop}%
\bibitem [{\citenamefont {Solà}\ \emph {et~al.}(2017)\citenamefont {Solà},
  \citenamefont {Gómez-Valent},\ and\ \citenamefont
  {de~Cruz~Pérez}}]{Sola:2017znb}%
  \BibitemOpen
  \bibfield  {author} {\bibinfo {author} {\bibfnamefont {J.}~\bibnamefont
  {Solà}}, \bibinfo {author} {\bibfnamefont {A.}~\bibnamefont
  {Gómez-Valent}}, \ and\ \bibinfo {author} {\bibfnamefont {J.}~\bibnamefont
  {de~Cruz~Pérez}},\ }\href {\doibase 10.1016/j.physletb.2017.09.073}
  {\bibfield  {journal} {\bibinfo  {journal} {Phys. Lett.}\ }\textbf {\bibinfo
  {volume} {B774}},\ \bibinfo {pages} {317} (\bibinfo {year} {2017})},\ \Eprint
  {http://arxiv.org/abs/1705.06723} {arXiv:1705.06723 [astro-ph.CO]}
  \BibitemShut {NoStop}%
\bibitem [{\citenamefont {di~Valentino~et
  al.}(2021{\natexlab{a}})}]{DIVALENTINO2021102605}%
  \BibitemOpen
  \bibfield  {author} {\bibinfo {author} {\bibfnamefont {E.}~\bibnamefont
  {di~Valentino~et al.}},\ }\href {\doibase
  https://doi.org/10.1016/j.astropartphys.2021.102605} {\bibfield  {journal}
  {\bibinfo  {journal} {Astroparticle Physics}\ }\textbf {\bibinfo {volume}
  {131}},\ \bibinfo {pages} {102605} (\bibinfo {year}
  {2021}{\natexlab{a}})}\BibitemShut {NoStop}%
\bibitem [{\citenamefont {di~Valentino~et
  al.}(2021{\natexlab{b}})}]{DIVALENTINO2021102604}%
  \BibitemOpen
  \bibfield  {author} {\bibinfo {author} {\bibfnamefont {E.}~\bibnamefont
  {di~Valentino~et al.}},\ }\href {\doibase
  https://doi.org/10.1016/j.astropartphys.2021.102604} {\bibfield  {journal}
  {\bibinfo  {journal} {Astroparticle Physics}\ }\textbf {\bibinfo {volume}
  {131}},\ \bibinfo {pages} {102604} (\bibinfo {year}
  {2021}{\natexlab{b}})}\BibitemShut {NoStop}%
\bibitem [{\citenamefont {Perivolaropoulos}\ and\ \citenamefont
  {Skara}(2021)}]{Perivolaropoulos:2021jda}%
  \BibitemOpen
  \bibfield  {author} {\bibinfo {author} {\bibfnamefont {L.}~\bibnamefont
  {Perivolaropoulos}}\ and\ \bibinfo {author} {\bibfnamefont {F.}~\bibnamefont
  {Skara}},\ }\href@noop {} {\  (\bibinfo {year} {2021})},\ \Eprint
  {http://arxiv.org/abs/2105.05208} {arXiv:2105.05208 [astro-ph.CO]}
  \BibitemShut {NoStop}%
\bibitem [{\citenamefont {Lusso}\ \emph {et~al.}(2019)\citenamefont {Lusso},
  \citenamefont {Piedipalumbo}, \citenamefont {Risaliti}, \citenamefont
  {Paolillo}, \citenamefont {Bisogni}, \citenamefont {Nardini},\ and\
  \citenamefont {Amati}}]{Lusso:2019akb}%
  \BibitemOpen
  \bibfield  {author} {\bibinfo {author} {\bibfnamefont {E.}~\bibnamefont
  {Lusso}}, \bibinfo {author} {\bibfnamefont {E.}~\bibnamefont {Piedipalumbo}},
  \bibinfo {author} {\bibfnamefont {G.}~\bibnamefont {Risaliti}}, \bibinfo
  {author} {\bibfnamefont {M.}~\bibnamefont {Paolillo}}, \bibinfo {author}
  {\bibfnamefont {S.}~\bibnamefont {Bisogni}}, \bibinfo {author} {\bibfnamefont
  {E.}~\bibnamefont {Nardini}}, \ and\ \bibinfo {author} {\bibfnamefont
  {L.}~\bibnamefont {Amati}},\ }\href {\doibase 10.1051/0004-6361/201936223}
  {\bibfield  {journal} {\bibinfo  {journal} {Astron. Astrophys.}\ }\textbf
  {\bibinfo {volume} {628}},\ \bibinfo {pages} {L4} (\bibinfo {year} {2019})},\
  \Eprint {http://arxiv.org/abs/1907.07692} {arXiv:1907.07692 [astro-ph.CO]}
  \BibitemShut {NoStop}%
\bibitem [{\citenamefont {Risaliti}\ and\ \citenamefont
  {Lusso}(2019)}]{Risaliti:2018reu}%
  \BibitemOpen
  \bibfield  {author} {\bibinfo {author} {\bibfnamefont {G.}~\bibnamefont
  {Risaliti}}\ and\ \bibinfo {author} {\bibfnamefont {E.}~\bibnamefont
  {Lusso}},\ }\href {\doibase 10.1038/s41550-018-0657-z} {\bibfield  {journal}
  {\bibinfo  {journal} {Nat. Astron.}\ }\textbf {\bibinfo {volume} {3}},\
  \bibinfo {pages} {272} (\bibinfo {year} {2019})},\ \Eprint
  {http://arxiv.org/abs/1811.02590} {arXiv:1811.02590 [astro-ph.CO]}
  \BibitemShut {NoStop}%
\bibitem [{\citenamefont {Lusso}\ \emph {et~al.}(2020)\citenamefont {Lusso},
  \citenamefont {Risaliti}, \citenamefont {Nardini}, \citenamefont
  {Bargiacchi}, \citenamefont {Benetti}, \citenamefont {Bisogni}, \citenamefont
  {Capozziello}, \citenamefont {Civano}, \citenamefont {Eggleston},
  \citenamefont {Elvis},\ and\ \citenamefont {et~al.}}]{Lusso_2020}%
  \BibitemOpen
  \bibfield  {author} {\bibinfo {author} {\bibfnamefont {E.}~\bibnamefont
  {Lusso}}, \bibinfo {author} {\bibfnamefont {G.}~\bibnamefont {Risaliti}},
  \bibinfo {author} {\bibfnamefont {E.}~\bibnamefont {Nardini}}, \bibinfo
  {author} {\bibfnamefont {G.}~\bibnamefont {Bargiacchi}}, \bibinfo {author}
  {\bibfnamefont {M.}~\bibnamefont {Benetti}}, \bibinfo {author} {\bibfnamefont
  {S.}~\bibnamefont {Bisogni}}, \bibinfo {author} {\bibfnamefont
  {S.}~\bibnamefont {Capozziello}}, \bibinfo {author} {\bibfnamefont
  {F.}~\bibnamefont {Civano}}, \bibinfo {author} {\bibfnamefont
  {L.}~\bibnamefont {Eggleston}}, \bibinfo {author} {\bibfnamefont
  {M.}~\bibnamefont {Elvis}}, \ and\ \bibinfo {author} {\bibnamefont
  {et~al.}},\ }\href {\doibase 10.1051/0004-6361/202038899} {\bibfield
  {journal} {\bibinfo  {journal} {\aap}\ }\textbf {\bibinfo {volume} {642}},\
  \bibinfo {pages} {A150} (\bibinfo {year} {2020})}\BibitemShut {NoStop}%
\bibitem [{\citenamefont {González-Morán}\ \emph {et~al.}(2021)\citenamefont
  {González-Morán}, \citenamefont {Chávez}, \citenamefont {Terlevich},
  \citenamefont {Terlevich}, \citenamefont {Fernández-Arenas}, \citenamefont
  {Bresolin}, \citenamefont {Plionis}, \citenamefont {Melnick}, \citenamefont
  {Basilakos},\ and\ \citenamefont {Telles}}]{Gonz_lez_Mor_n_2021}%
  \BibitemOpen
  \bibfield  {author} {\bibinfo {author} {\bibfnamefont {A.~L.}\ \bibnamefont
  {González-Morán}}, \bibinfo {author} {\bibfnamefont {R.}~\bibnamefont
  {Chávez}}, \bibinfo {author} {\bibfnamefont {E.}~\bibnamefont {Terlevich}},
  \bibinfo {author} {\bibfnamefont {R.}~\bibnamefont {Terlevich}}, \bibinfo
  {author} {\bibfnamefont {D.}~\bibnamefont {Fernández-Arenas}}, \bibinfo
  {author} {\bibfnamefont {F.}~\bibnamefont {Bresolin}}, \bibinfo {author}
  {\bibfnamefont {M.}~\bibnamefont {Plionis}}, \bibinfo {author} {\bibfnamefont
  {J.}~\bibnamefont {Melnick}}, \bibinfo {author} {\bibfnamefont
  {S.}~\bibnamefont {Basilakos}}, \ and\ \bibinfo {author} {\bibfnamefont
  {E.}~\bibnamefont {Telles}},\ }\href {\doibase 10.1093/mnras/stab1385}
  {\bibfield  {journal} {\bibinfo  {journal} {Monthly Notices of the Royal
  Astronomical Society}\ } (\bibinfo {year} {2021}),\
  10.1093/mnras/stab1385}\BibitemShut {NoStop}%
\bibitem [{\citenamefont {Gonz\'alez-Mor\'an}\ \emph
  {et~al.}(2019)\citenamefont {Gonz\'alez-Mor\'an}, \citenamefont {Ch\'avez},
  \citenamefont {Terlevich}, \citenamefont {Terlevich}, \citenamefont
  {Bresolin}, \citenamefont {Fern\'andez-Arenas}, \citenamefont {Plionis},
  \citenamefont {Basilakos}, \citenamefont {Melnick},\ and\ \citenamefont
  {Telles}}]{Gonzalez-Moran:2019uij}%
  \BibitemOpen
  \bibfield  {author} {\bibinfo {author} {\bibfnamefont {A.~L.}\ \bibnamefont
  {Gonz\'alez-Mor\'an}}, \bibinfo {author} {\bibfnamefont {R.}~\bibnamefont
  {Ch\'avez}}, \bibinfo {author} {\bibfnamefont {R.}~\bibnamefont {Terlevich}},
  \bibinfo {author} {\bibfnamefont {E.}~\bibnamefont {Terlevich}}, \bibinfo
  {author} {\bibfnamefont {F.}~\bibnamefont {Bresolin}}, \bibinfo {author}
  {\bibfnamefont {D.}~\bibnamefont {Fern\'andez-Arenas}}, \bibinfo {author}
  {\bibfnamefont {M.}~\bibnamefont {Plionis}}, \bibinfo {author} {\bibfnamefont
  {S.}~\bibnamefont {Basilakos}}, \bibinfo {author} {\bibfnamefont
  {J.}~\bibnamefont {Melnick}}, \ and\ \bibinfo {author} {\bibfnamefont
  {E.}~\bibnamefont {Telles}},\ }\href {\doibase 10.1093/mnras/stz1577}
  {\bibfield  {journal} {\bibinfo  {journal} {Mon. Not. Roy. Astron. Soc.}\
  }\textbf {\bibinfo {volume} {487}},\ \bibinfo {pages} {4669} (\bibinfo {year}
  {2019})},\ \Eprint {http://arxiv.org/abs/1906.02195} {arXiv:1906.02195
  [astro-ph.GA]} \BibitemShut {NoStop}%
\bibitem [{\citenamefont {Terlevich}\ \emph {et~al.}(2015)\citenamefont
  {Terlevich}, \citenamefont {Terlevich}, \citenamefont {Melnick},
  \citenamefont {Ch\'avez}, \citenamefont {Plionis}, \citenamefont {Bresolin},\
  and\ \citenamefont {Basilakos}}]{Terlevich:2015toa}%
  \BibitemOpen
  \bibfield  {author} {\bibinfo {author} {\bibfnamefont {R.}~\bibnamefont
  {Terlevich}}, \bibinfo {author} {\bibfnamefont {E.}~\bibnamefont
  {Terlevich}}, \bibinfo {author} {\bibfnamefont {J.}~\bibnamefont {Melnick}},
  \bibinfo {author} {\bibfnamefont {R.}~\bibnamefont {Ch\'avez}}, \bibinfo
  {author} {\bibfnamefont {M.}~\bibnamefont {Plionis}}, \bibinfo {author}
  {\bibfnamefont {F.}~\bibnamefont {Bresolin}}, \ and\ \bibinfo {author}
  {\bibfnamefont {S.}~\bibnamefont {Basilakos}},\ }\href {\doibase
  10.1093/mnras/stv1128} {\bibfield  {journal} {\bibinfo  {journal} {Mon. Not.
  Roy. Astron. Soc.}\ }\textbf {\bibinfo {volume} {451}},\ \bibinfo {pages}
  {3001} (\bibinfo {year} {2015})},\ \Eprint {http://arxiv.org/abs/1505.04376}
  {arXiv:1505.04376 [astro-ph.CO]} \BibitemShut {NoStop}%
\bibitem [{\citenamefont {Scolnic}\ \emph {et~al.}(2018)\citenamefont {Scolnic}
  \emph {et~al.}}]{Scolnic:2017caz}%
  \BibitemOpen
  \bibfield  {author} {\bibinfo {author} {\bibfnamefont {D.~M.}\ \bibnamefont
  {Scolnic}} \emph {et~al.},\ }\href {\doibase 10.3847/1538-4357/aab9bb}
  {\bibfield  {journal} {\bibinfo  {journal} {Astrophys. J.}\ }\textbf
  {\bibinfo {volume} {859}},\ \bibinfo {pages} {101} (\bibinfo {year}
  {2018})},\ \Eprint {http://arxiv.org/abs/1710.00845} {arXiv:1710.00845
  [astro-ph.CO]} \BibitemShut {NoStop}%
\bibitem [{\citenamefont {Liao}\ \emph {et~al.}(2019)\citenamefont {Liao},
  \citenamefont {Shafieloo}, \citenamefont {Keeley},\ and\ \citenamefont
  {Linder}}]{Liao:2019qoc}%
  \BibitemOpen
  \bibfield  {author} {\bibinfo {author} {\bibfnamefont {K.}~\bibnamefont
  {Liao}}, \bibinfo {author} {\bibfnamefont {A.}~\bibnamefont {Shafieloo}},
  \bibinfo {author} {\bibfnamefont {R.~E.}\ \bibnamefont {Keeley}}, \ and\
  \bibinfo {author} {\bibfnamefont {E.~V.}\ \bibnamefont {Linder}},\
  }\href@noop {} {\  (\bibinfo {year} {2019})},\ \Eprint
  {http://arxiv.org/abs/1908.04967} {arXiv:1908.04967 [astro-ph.CO]}
  \BibitemShut {NoStop}%
\bibitem [{\citenamefont {Zhang}\ and\ \citenamefont
  {Li}(2018)}]{Zhang:2018gjb}%
  \BibitemOpen
  \bibfield  {author} {\bibinfo {author} {\bibfnamefont {M.-J.}\ \bibnamefont
  {Zhang}}\ and\ \bibinfo {author} {\bibfnamefont {H.}~\bibnamefont {Li}},\
  }\href {\doibase 10.1140/epjc/s10052-018-5953-3} {\bibfield  {journal}
  {\bibinfo  {journal} {Eur. Phys. J.}\ }\textbf {\bibinfo {volume} {C78}},\
  \bibinfo {pages} {460} (\bibinfo {year} {2018})},\ \Eprint
  {http://arxiv.org/abs/1806.02981} {arXiv:1806.02981 [astro-ph.CO]}
  \BibitemShut {NoStop}%
\bibitem [{\citenamefont {Gómez-Valent}\ and\ \citenamefont
  {Amendola}(2018)}]{Gomez-Valent:2018hwc}%
  \BibitemOpen
  \bibfield  {author} {\bibinfo {author} {\bibfnamefont {A.}~\bibnamefont
  {Gómez-Valent}}\ and\ \bibinfo {author} {\bibfnamefont {L.}~\bibnamefont
  {Amendola}},\ }\href {\doibase 10.1088/1475-7516/2018/04/051} {\bibfield
  {journal} {\bibinfo  {journal} {JCAP}\ }\textbf {\bibinfo {volume} {1804}},\
  \bibinfo {pages} {051} (\bibinfo {year} {2018})},\ \Eprint
  {http://arxiv.org/abs/1802.01505} {arXiv:1802.01505 [astro-ph.CO]}
  \BibitemShut {NoStop}%
\bibitem [{\citenamefont {Melia}\ and\ \citenamefont
  {Yennapureddy}(2018)}]{Melia:2018tzi}%
  \BibitemOpen
  \bibfield  {author} {\bibinfo {author} {\bibfnamefont {F.}~\bibnamefont
  {Melia}}\ and\ \bibinfo {author} {\bibfnamefont {M.~K.}\ \bibnamefont
  {Yennapureddy}},\ }\href {\doibase 10.1088/1475-7516/2018/02/034} {\bibfield
  {journal} {\bibinfo  {journal} {JCAP}\ }\textbf {\bibinfo {volume} {1802}},\
  \bibinfo {pages} {034} (\bibinfo {year} {2018})},\ \Eprint
  {http://arxiv.org/abs/1802.02255} {arXiv:1802.02255 [astro-ph.CO]}
  \BibitemShut {NoStop}%
\bibitem [{\citenamefont {Mehrabi}\ and\ \citenamefont
  {Basilakos}(2020)}]{Mehrabi_2020}%
  \BibitemOpen
  \bibfield  {author} {\bibinfo {author} {\bibfnamefont {A.}~\bibnamefont
  {Mehrabi}}\ and\ \bibinfo {author} {\bibfnamefont {S.}~\bibnamefont
  {Basilakos}},\ }\href {\doibase 10.1140/epjc/s10052-020-8221-2} {\bibfield
  {journal} {\bibinfo  {journal} {The European Physical Journal C}\ }\textbf
  {\bibinfo {volume} {80}} (\bibinfo {year} {2020}),\
  10.1140/epjc/s10052-020-8221-2}\BibitemShut {NoStop}%
\bibitem [{\citenamefont {Fern\'andez~Arenas}\ \emph
  {et~al.}(2018)\citenamefont {Fern\'andez~Arenas}, \citenamefont {Terlevich},
  \citenamefont {Terlevich}, \citenamefont {Melnick}, \citenamefont {Ch\'avez},
  \citenamefont {Bresolin}, \citenamefont {Telles}, \citenamefont {Plionis},\
  and\ \citenamefont {Basilakos}}]{Fernandez-Arenas:2017isq}%
  \BibitemOpen
  \bibfield  {author} {\bibinfo {author} {\bibfnamefont {D.}~\bibnamefont
  {Fern\'andez~Arenas}}, \bibinfo {author} {\bibfnamefont {E.}~\bibnamefont
  {Terlevich}}, \bibinfo {author} {\bibfnamefont {R.}~\bibnamefont
  {Terlevich}}, \bibinfo {author} {\bibfnamefont {J.}~\bibnamefont {Melnick}},
  \bibinfo {author} {\bibfnamefont {R.}~\bibnamefont {Ch\'avez}}, \bibinfo
  {author} {\bibfnamefont {F.}~\bibnamefont {Bresolin}}, \bibinfo {author}
  {\bibfnamefont {E.}~\bibnamefont {Telles}}, \bibinfo {author} {\bibfnamefont
  {M.}~\bibnamefont {Plionis}}, \ and\ \bibinfo {author} {\bibfnamefont
  {S.}~\bibnamefont {Basilakos}},\ }\href {\doibase 10.1093/mnras/stx2710}
  {\bibfield  {journal} {\bibinfo  {journal} {Mon. Not. Roy. Astron. Soc.}\
  }\textbf {\bibinfo {volume} {474}},\ \bibinfo {pages} {1250} (\bibinfo {year}
  {2018})},\ \Eprint {http://arxiv.org/abs/1710.05951} {arXiv:1710.05951
  [astro-ph.CO]} \BibitemShut {NoStop}%
\bibitem [{\citenamefont {Ch\'avez}\ \emph {et~al.}(2016)\citenamefont
  {Ch\'avez}, \citenamefont {Plionis}, \citenamefont {Basilakos}, \citenamefont
  {Terlevich}, \citenamefont {Terlevich}, \citenamefont {Melnick},
  \citenamefont {Bresolin},\ and\ \citenamefont
  {Gonz\'alez-Mor\'an}}]{Chavez:2016epc}%
  \BibitemOpen
  \bibfield  {author} {\bibinfo {author} {\bibfnamefont {R.}~\bibnamefont
  {Ch\'avez}}, \bibinfo {author} {\bibfnamefont {M.}~\bibnamefont {Plionis}},
  \bibinfo {author} {\bibfnamefont {S.}~\bibnamefont {Basilakos}}, \bibinfo
  {author} {\bibfnamefont {R.}~\bibnamefont {Terlevich}}, \bibinfo {author}
  {\bibfnamefont {E.}~\bibnamefont {Terlevich}}, \bibinfo {author}
  {\bibfnamefont {J.}~\bibnamefont {Melnick}}, \bibinfo {author} {\bibfnamefont
  {F.}~\bibnamefont {Bresolin}}, \ and\ \bibinfo {author} {\bibfnamefont
  {A.~L.}\ \bibnamefont {Gonz\'alez-Mor\'an}},\ }\href {\doibase
  10.1093/mnras/stw1813} {\bibfield  {journal} {\bibinfo  {journal} {Mon. Not.
  Roy. Astron. Soc.}\ }\textbf {\bibinfo {volume} {462}},\ \bibinfo {pages}
  {2431} (\bibinfo {year} {2016})},\ \Eprint {http://arxiv.org/abs/1607.06458}
  {arXiv:1607.06458 [astro-ph.CO]} \BibitemShut {NoStop}%
\bibitem [{\citenamefont {Plionis}\ \emph {et~al.}(2010)\citenamefont
  {Plionis}, \citenamefont {Terlevich}, \citenamefont {Basilakos},
  \citenamefont {Bresolin}, \citenamefont {Terlevich}, \citenamefont
  {Melnick},\ and\ \citenamefont {Chavez}}]{Plionis:2009wp}%
  \BibitemOpen
  \bibfield  {author} {\bibinfo {author} {\bibfnamefont {M.}~\bibnamefont
  {Plionis}}, \bibinfo {author} {\bibfnamefont {R.}~\bibnamefont {Terlevich}},
  \bibinfo {author} {\bibfnamefont {S.}~\bibnamefont {Basilakos}}, \bibinfo
  {author} {\bibfnamefont {F.}~\bibnamefont {Bresolin}}, \bibinfo {author}
  {\bibfnamefont {E.}~\bibnamefont {Terlevich}}, \bibinfo {author}
  {\bibfnamefont {J.}~\bibnamefont {Melnick}}, \ and\ \bibinfo {author}
  {\bibfnamefont {R.}~\bibnamefont {Chavez}},\ }\href {\doibase
  10.1063/1.3462645} {\bibfield  {journal} {\bibinfo  {journal} {AIP Conf.
  Proc.}\ }\textbf {\bibinfo {volume} {1241}},\ \bibinfo {pages} {267}
  (\bibinfo {year} {2010})},\ \Eprint {http://arxiv.org/abs/0911.3198}
  {arXiv:0911.3198 [astro-ph.CO]} \BibitemShut {NoStop}%
\bibitem [{\citenamefont {Plionis}\ \emph {et~al.}(2011)\citenamefont
  {Plionis}, \citenamefont {Terlevich}, \citenamefont {Basilakos},
  \citenamefont {Bresolin}, \citenamefont {Terlevich}, \citenamefont
  {Melnick},\ and\ \citenamefont {Chavez}}]{Plionis:2011jj}%
  \BibitemOpen
  \bibfield  {author} {\bibinfo {author} {\bibfnamefont {M.}~\bibnamefont
  {Plionis}}, \bibinfo {author} {\bibfnamefont {R.}~\bibnamefont {Terlevich}},
  \bibinfo {author} {\bibfnamefont {S.}~\bibnamefont {Basilakos}}, \bibinfo
  {author} {\bibfnamefont {F.}~\bibnamefont {Bresolin}}, \bibinfo {author}
  {\bibfnamefont {E.}~\bibnamefont {Terlevich}}, \bibinfo {author}
  {\bibfnamefont {J.}~\bibnamefont {Melnick}}, \ and\ \bibinfo {author}
  {\bibfnamefont {R.}~\bibnamefont {Chavez}},\ }\href {\doibase
  10.1111/j.1365-2966.2011.19247.x} {\bibfield  {journal} {\bibinfo  {journal}
  {Mon. Not. Roy. Astron. Soc.}\ }\textbf {\bibinfo {volume} {416}},\ \bibinfo
  {pages} {2981} (\bibinfo {year} {2011})},\ \Eprint
  {http://arxiv.org/abs/1106.4558} {arXiv:1106.4558 [astro-ph.CO]} \BibitemShut
  {NoStop}%
\bibitem [{\citenamefont {Plionis}\ \emph {et~al.}(2009)\citenamefont
  {Plionis}, \citenamefont {Terlevich}, \citenamefont {Basilakos},
  \citenamefont {Bresolin}, \citenamefont {Terlevich}, \citenamefont
  {Melnick},\ and\ \citenamefont {Georgantopoulos}}]{Plionis:2009up}%
  \BibitemOpen
  \bibfield  {author} {\bibinfo {author} {\bibfnamefont {M.}~\bibnamefont
  {Plionis}}, \bibinfo {author} {\bibfnamefont {R.}~\bibnamefont {Terlevich}},
  \bibinfo {author} {\bibfnamefont {S.}~\bibnamefont {Basilakos}}, \bibinfo
  {author} {\bibfnamefont {F.}~\bibnamefont {Bresolin}}, \bibinfo {author}
  {\bibfnamefont {E.}~\bibnamefont {Terlevich}}, \bibinfo {author}
  {\bibfnamefont {J.}~\bibnamefont {Melnick}}, \ and\ \bibinfo {author}
  {\bibfnamefont {I.}~\bibnamefont {Georgantopoulos}},\ }\href {\doibase
  10.1088/1742-6596/189/1/012032} {\bibfield  {journal} {\bibinfo  {journal}
  {J. Phys. Conf. Ser.}\ }\textbf {\bibinfo {volume} {189}},\ \bibinfo {pages}
  {012032} (\bibinfo {year} {2009})},\ \Eprint {http://arxiv.org/abs/0903.0131}
  {arXiv:0903.0131 [astro-ph.CO]} \BibitemShut {NoStop}%
\bibitem [{\citenamefont {Melnick}\ \emph {et~al.}(2000)\citenamefont
  {Melnick}, \citenamefont {Terlevich},\ and\ \citenamefont
  {Terlevich}}]{Melnick:1999qb}%
  \BibitemOpen
  \bibfield  {author} {\bibinfo {author} {\bibfnamefont {J.}~\bibnamefont
  {Melnick}}, \bibinfo {author} {\bibfnamefont {R.}~\bibnamefont {Terlevich}},
  \ and\ \bibinfo {author} {\bibfnamefont {E.}~\bibnamefont {Terlevich}},\
  }\href {\doibase 10.1046/j.1365-8711.2000.03112.x} {\bibfield  {journal}
  {\bibinfo  {journal} {Mon. Not. Roy. Astron. Soc.}\ }\textbf {\bibinfo
  {volume} {311}},\ \bibinfo {pages} {629} (\bibinfo {year} {2000})},\ \Eprint
  {http://arxiv.org/abs/astro-ph/9908346} {arXiv:astro-ph/9908346} \BibitemShut
  {NoStop}%
\bibitem [{Note1()}]{Note1}%
  \BibitemOpen
  \bibinfo {note} {For the concordance $\Lambda $CDM model we have $q_{\Lambda
  }(z)=\protect \frac {3}{2}\Omega _{m}(z)-1$ and $j_{\Lambda }(z)=1$, where
  $\Omega _{m}(z)=\Omega _{m0}(1+z)^{3}/E^{2}_{\Lambda }(z)$ with
  $E^{2}_{\Lambda }(z)=H^{2}_{\Lambda }(z)/H^{2}_{0}=\Omega
  _{m0}(1+z)^{3}+1-\Omega _{m0}$.}\BibitemShut {Stop}%
\bibitem [{\citenamefont {Rasmussen}\ and\ \citenamefont
  {Williams}(2005)}]{10.5555/1162254}%
  \BibitemOpen
  \bibfield  {author} {\bibinfo {author} {\bibfnamefont {C.~E.}\ \bibnamefont
  {Rasmussen}}\ and\ \bibinfo {author} {\bibfnamefont {C.~K.~I.}\ \bibnamefont
  {Williams}},\ }\href@noop {} {\emph {\bibinfo {title} {Gaussian Processes for
  Machine Learning (Adaptive Computation and Machine Learning)}}}\ (\bibinfo
  {publisher} {The MIT Press},\ \bibinfo {year} {2005})\BibitemShut {NoStop}%
\bibitem [{\citenamefont {Rasmussen}\ and\ \citenamefont
  {Williams}(2006)}]{GP_book}%
  \BibitemOpen
  \bibfield  {author} {\bibinfo {author} {\bibfnamefont {C.}~\bibnamefont
  {Rasmussen}}\ and\ \bibinfo {author} {\bibfnamefont {C.}~\bibnamefont
  {Williams}},\ }\href@noop {} {\emph {\bibinfo {title} {Gaussian Processes for
  Machine Learning}}},\ Adaptive Computation and Machine Learning\ (\bibinfo
  {publisher} {MIT Press},\ \bibinfo {address} {Cambridge, MA, USA},\ \bibinfo
  {year} {2006})\ p.\ \bibinfo {pages} {248}\BibitemShut {NoStop}%
\bibitem [{\citenamefont {Seikel}\ \emph {et~al.}(2012)\citenamefont {Seikel},
  \citenamefont {Clarkson},\ and\ \citenamefont {Smith}}]{Seikel:2012uu}%
  \BibitemOpen
  \bibfield  {author} {\bibinfo {author} {\bibfnamefont {M.}~\bibnamefont
  {Seikel}}, \bibinfo {author} {\bibfnamefont {C.}~\bibnamefont {Clarkson}}, \
  and\ \bibinfo {author} {\bibfnamefont {M.}~\bibnamefont {Smith}},\ }\href
  {\doibase 10.1088/1475-7516/2012/06/036} {\bibfield  {journal} {\bibinfo
  {journal} {JCAP}\ }\textbf {\bibinfo {volume} {1206}},\ \bibinfo {pages}
  {036} (\bibinfo {year} {2012})},\ \Eprint {http://arxiv.org/abs/1204.2832}
  {arXiv:1204.2832 [astro-ph.CO]} \BibitemShut {NoStop}%
\bibitem [{\citenamefont {Pedregosa}\ \emph {et~al.}(2011)\citenamefont
  {Pedregosa}, \citenamefont {Varoquaux}, \citenamefont {Gramfort},
  \citenamefont {Michel}, \citenamefont {Thirion}, \citenamefont {Grisel},
  \citenamefont {Blondel}, \citenamefont {Prettenhofer}, \citenamefont {Weiss},
  \citenamefont {Dubourg}, \citenamefont {Vanderplas}, \citenamefont {Passos},
  \citenamefont {Cournapeau}, \citenamefont {Brucher}, \citenamefont {Perrot},\
  and\ \citenamefont {Duchesnay}}]{scikit-learn}%
  \BibitemOpen
  \bibfield  {author} {\bibinfo {author} {\bibfnamefont {F.}~\bibnamefont
  {Pedregosa}}, \bibinfo {author} {\bibfnamefont {G.}~\bibnamefont
  {Varoquaux}}, \bibinfo {author} {\bibfnamefont {A.}~\bibnamefont {Gramfort}},
  \bibinfo {author} {\bibfnamefont {V.}~\bibnamefont {Michel}}, \bibinfo
  {author} {\bibfnamefont {B.}~\bibnamefont {Thirion}}, \bibinfo {author}
  {\bibfnamefont {O.}~\bibnamefont {Grisel}}, \bibinfo {author} {\bibfnamefont
  {M.}~\bibnamefont {Blondel}}, \bibinfo {author} {\bibfnamefont
  {P.}~\bibnamefont {Prettenhofer}}, \bibinfo {author} {\bibfnamefont
  {R.}~\bibnamefont {Weiss}}, \bibinfo {author} {\bibfnamefont
  {V.}~\bibnamefont {Dubourg}}, \bibinfo {author} {\bibfnamefont
  {J.}~\bibnamefont {Vanderplas}}, \bibinfo {author} {\bibfnamefont
  {A.}~\bibnamefont {Passos}}, \bibinfo {author} {\bibfnamefont
  {D.}~\bibnamefont {Cournapeau}}, \bibinfo {author} {\bibfnamefont
  {M.}~\bibnamefont {Brucher}}, \bibinfo {author} {\bibfnamefont
  {M.}~\bibnamefont {Perrot}}, \ and\ \bibinfo {author} {\bibfnamefont
  {E.}~\bibnamefont {Duchesnay}},\ }\href@noop {} {\bibfield  {journal}
  {\bibinfo  {journal} {Journal of Machine Learning Research}\ }\textbf
  {\bibinfo {volume} {12}},\ \bibinfo {pages} {2825} (\bibinfo {year}
  {2011})}\BibitemShut {NoStop}%
\bibitem [{\citenamefont {Melnick}\ \emph {et~al.}(1988)\citenamefont
  {Melnick}, \citenamefont {Terlevich},\ and\ \citenamefont
  {Moles}}]{10.1093/mnras/235.1.297}%
  \BibitemOpen
  \bibfield  {author} {\bibinfo {author} {\bibfnamefont {J.}~\bibnamefont
  {Melnick}}, \bibinfo {author} {\bibfnamefont {R.}~\bibnamefont {Terlevich}},
  \ and\ \bibinfo {author} {\bibfnamefont {M.}~\bibnamefont {Moles}},\ }\href
  {\doibase 10.1093/mnras/235.1.297} {\bibfield  {journal} {\bibinfo  {journal}
  {Monthly Notices of the Royal Astronomical Society}\ }\textbf {\bibinfo
  {volume} {235}},\ \bibinfo {pages} {297} (\bibinfo {year} {1988})},\ \Eprint
  {http://arxiv.org/abs/https://academic.oup.com/mnras/article-pdf/235/1/297/3145334/mnras235-0297.pdf}
  {https://academic.oup.com/mnras/article-pdf/235/1/297/3145334/mnras235-0297.pdf}
  \BibitemShut {NoStop}%
\bibitem [{\citenamefont {Wei}\ \emph {et~al.}(2016)\citenamefont {Wei},
  \citenamefont {Wu},\ and\ \citenamefont {Melia}}]{Wei:2016jqa}%
  \BibitemOpen
  \bibfield  {author} {\bibinfo {author} {\bibfnamefont {J.-J.}\ \bibnamefont
  {Wei}}, \bibinfo {author} {\bibfnamefont {X.-F.}\ \bibnamefont {Wu}}, \ and\
  \bibinfo {author} {\bibfnamefont {F.}~\bibnamefont {Melia}},\ }\href
  {\doibase 10.1093/mnras/stw2057} {\bibfield  {journal} {\bibinfo  {journal}
  {Mon. Not. Roy. Astron. Soc.}\ }\textbf {\bibinfo {volume} {463}},\ \bibinfo
  {pages} {1144} (\bibinfo {year} {2016})},\ \Eprint
  {http://arxiv.org/abs/1608.02070} {arXiv:1608.02070 [astro-ph.CO]}
  \BibitemShut {NoStop}%
\bibitem [{\citenamefont {Yennapureddy}\ and\ \citenamefont
  {Melia}(2017)}]{Yennapureddy:2017vvb}%
  \BibitemOpen
  \bibfield  {author} {\bibinfo {author} {\bibfnamefont {M.~K.}\ \bibnamefont
  {Yennapureddy}}\ and\ \bibinfo {author} {\bibfnamefont {F.}~\bibnamefont
  {Melia}},\ }\href {\doibase 10.1088/1475-7516/2017/11/029} {\bibfield
  {journal} {\bibinfo  {journal} {JCAP}\ }\textbf {\bibinfo {volume} {11}},\
  \bibinfo {pages} {029} (\bibinfo {year} {2017})},\ \Eprint
  {http://arxiv.org/abs/1711.03454} {arXiv:1711.03454 [astro-ph.CO]}
  \BibitemShut {NoStop}%
\bibitem [{\citenamefont {Cao}\ \emph {et~al.}(2021)\citenamefont {Cao},
  \citenamefont {Ryan},\ and\ \citenamefont {Ratra}}]{Cao:2021cix}%
  \BibitemOpen
  \bibfield  {author} {\bibinfo {author} {\bibfnamefont {S.}~\bibnamefont
  {Cao}}, \bibinfo {author} {\bibfnamefont {J.}~\bibnamefont {Ryan}}, \ and\
  \bibinfo {author} {\bibfnamefont {B.}~\bibnamefont {Ratra}},\ }\href@noop {}
  {\  (\bibinfo {year} {2021})},\ \Eprint {http://arxiv.org/abs/2109.01987}
  {arXiv:2109.01987 [astro-ph.CO]} \BibitemShut {NoStop}%
\end{thebibliography}%

\end{document}